\documentclass[compsoc, conference, a4paper, 10pt, times, anonymous]{IEEEtran}

\usepackage{enumitem}
\usepackage{graphicx,subfigure}
\usepackage{caption}
\usepackage{import}
\usepackage{comment}
\usepackage{tikz}
\usepackage{amsmath}
\usepackage{MnSymbol}
\usepackage{booktabs}
\usepackage[linesnumbered,ruled,vlined]{algorithm2e}
\SetKwInput{KwInput}{Input}               
\SetKwInput{KwOutput}{Output}
\usepackage{xurl}
\usepackage{soul}
\usepackage{multirow}
\usepackage{tablefootnote}
\usepackage[hidelinks]{hyperref}

\begin{document}

\title{Dynamic Frequency-Based Fingerprinting Attacks against Modern Sandbox Environments}

\author{Anonymous Author(s)}

\begin{comment}
% Submissions should be anonymized. See the CFP for details on how to anonymize your paper, including any references to your own work.
%\author{\em Anonymous Authors}

% The author information is skipped here, but can be used to include author information in the publication.
\iffalse
\author{\IEEEauthorblockN{1\textsuperscript{st} Given Names Surname}
\IEEEauthorblockA{\textit{Affiliation} \\
City, Country \\
email address or website URL}
\and
\IEEEauthorblockN{2\textsuperscript{nd} Given Names Surname}
\IEEEauthorblockA{\textit{Affiliation} \\
City, Country \\
email address or website URL}
\and
\IEEEauthorblockN{3\textsuperscript{rd} Given Names Surname}
\IEEEauthorblockA{\textit{Affiliation} \\
City, Country \\
email address or website URL}
%% IEEE format can accommodate up to six authors this way
}
\fi

\author{
  \IEEEauthorblockN{Author 1\IEEEauthorrefmark{1}, Author 2\IEEEauthorrefmark{2}, Author 3\IEEEauthorrefmark{3}}\\
  \IEEEauthorblockA{\IEEEauthorrefmark{1}Affiliation 1}\\
  \IEEEauthorblockA{\IEEEauthorrefmark{2}Affiliation 2}\\
  \IEEEauthorblockA{\IEEEauthorrefmark{3}Affiliation 3}
  \and
  \IEEEauthorblockN{Author 4\IEEEauthorrefmark{4}, Author 5\IEEEauthorrefmark{5}}\\
  \IEEEauthorblockA{\IEEEauthorrefmark{4}Affiliation 4}\\
  \IEEEauthorblockA{\IEEEauthorrefmark{5}Affiliation 5}
}
\end{comment}
\makeatletter
\renewcommand\IEEEauthorrefmark[1]{\textsuperscript{#1}}
\makeatother

\author{\IEEEauthorblockN{Debopriya Roy Dipta}%\IEEEauthorrefmark{1}}
\IEEEauthorblockA{\textit{Iowa State University} \\
Ames, IA, USA \\
roydipta@iastate.edu
}

\\
\IEEEauthorblockN{Eduard Marin Fabregas}%\IEEEauthorrefmark{4}}
\IEEEauthorblockA{\textit{Telefonica Research} \\
Barcelona, Spain \\
eduard.marinfabregas@telefonica.com
}

\and
\IEEEauthorblockN{Thore Tiemann}%\IEEEauthorrefmark{2}}
\IEEEauthorblockA{\textit{University of Lüebeck} \\
Lübeck, Germany \\
t.tiemann@uni-luebeck.de
}
\and
\IEEEauthorblockN{Berk Gulmezoglu}%\IEEEauthorrefmark{3}}
\IEEEauthorblockA{\textit{Iowa State University} \\
Ames, IA, USA \\
bgulmez@iastate.edu
}
\\
\IEEEauthorblockN{Thomas Eisenbarth}%\IEEEauthorrefmark{5}}
\IEEEauthorblockA{\textit{University of Lüebeck} \\
Lübeck, Germany \\
thomas.eisenbarth@uni-luebeck.de
}
}

\maketitle
%\pagenumbering{arabic}

\begin{abstract}
The cloud computing landscape has evolved significantly in recent years, embracing various sandboxes to meet the diverse demands of modern cloud applications. These sandboxes encompass container-based technologies like Docker and gVisor, microVM-based solutions like Firecracker, and security-centric sandboxes relying on Trusted Execution Environments (TEEs) such as Intel SGX and AMD SEV. However, the practice of placing multiple tenants on shared physical hardware raises security and privacy concerns, most notably side-channel attacks. %Malicious entities in the cloud environment with only user-space privileges can still monitor tenants' activity and compromise their privacy and security stealthily.

In this paper, we investigate the possibility of fingerprinting containers through CPU frequency reporting sensors in Intel and AMD CPUs. One key enabler of our attack is that the current CPU frequency information can be accessed by user-space attackers.
We demonstrate that Docker images exhibit a unique frequency signature, enabling the distinction of different containers with up to 84.5\% accuracy even when multiple containers are running simultaneously in different cores.
Additionally, we assess the effectiveness of our attack when performed against several sandboxes deployed in cloud environments, including Google's gVisor, AWS' Firecracker, and TEE-based platforms like Gramine (utilizing Intel SGX) and AMD SEV.
Our empirical results show that these attacks can also be carried out successfully against all of these sandboxes in less than 40 seconds, with an accuracy of over 70\% in all cases. 
%Our attack can be performed successfully in less than 30 seconds by leveraging the current CPU frequency information accessed by user-space attackers. in less than 30 seconds
%Finally, we sketch out a possible countermeasure using system call pattern monitoring to detect malicious activities without false positives and with a minimal performance overhead of 1.8\%.
Finally, we propose a noise injection-based  countermeasure to mitigate the proposed attack on cloud environments.

\end{abstract}
\pagenumbering{arabic}
\setcounter{page}{1}

\section{Introduction} 

%Each microservice encapsulates a specific business capability or function and can be developed, updated, deployed, and scaled independently.

In recent years, there has been a significant surge in the adoption of novel application models for developing applications in the cloud. These models are primarily designed to offer enhanced flexibility and agility, improved fault isolation, greater cost-efficiency, and more efficient resource utilization. One notable development is the emergence of the microservices paradigm as an alternative to the conventional practice of running monolithic applications within virtual machines (VMs). In the microservices model, an application is decomposed into a collection of small, autonomous, and loosely interconnected components (i.\,e., the microservices) that communicate with each other through well-defined APIs. Simultaneously, there has been a sudden increase in the demand for highly secure and privacy-focused applications in the cloud, driving the need for confidential computing solutions based on Trusted Execution Environments (TEEs). Inevitably, these significant developments also require the creation of new sandbox environments designed to meet the performance and security needs of modern cloud applications.

%Google's gVisor~\cite{gvisor} serves as a prime example in the former category, while Amazon's Firecracker~\cite{agache2020firecracker} stands as a representative example of the latter class of execution environments.

Container technologies like Docker~\cite{docker} emerged as a compelling alternative to traditional VMs, primarily due to their ability to address inherent VM limitations such as slow start-up times and resource-intensive nature. Docker containers represented a significant step forward in the development and management of cloud applications, albeit with an important trade-off in security~\cite{7742298}. In light of this, major cloud service providers chose to develop their own execution environments, broadly categorized into two main approaches: (i) those built upon container technologies (e.g., Google's gVisor~\cite{gvisor}) and (ii) those focused on developing lightweight virtual machines (e.g., Amazon's Firecracker~\cite{agache2020firecracker}). Within the context of confidential computing, the landscape has similarly evolved to encompass two distinct styles for protecting cloud applications. On the one hand, new TEEs, such as AMD Secure Encrypted Virtualization (SEV)~\cite{sev2017seves}, have been introduced, enabling the protection of entire virtual machines. On the other hand, multiple frameworks~\cite{arnautov2016scone,gramine} have been proposed to ease the development of containers backed by a TEE like Intel Software Guard Extensions (SGX)~\cite{costan2016intel}. All in all, today’s cloud applications can run in various execution environments depending on the chosen cloud provider and the application’s security requirements.

Regardless of the execution environment or infrastructure employed, cloud providers frequently enhance server efficiency by allocating multiple tenants to the same physical hardware. This practice, known as co-location~\cite{wang2018peeking,ristenpart2009hey,inci2016co}, results in the sharing of various hardware components and OS features among these tenants. Unfortunately, the shared nature of hardware presents security and privacy risks for cloud computing users, as it can be susceptible to side-channel attacks~\cite{zhang2014cross,inci2015seriously}.
By exploiting shared hardware components, such as data/instruction cache~\cite{aciiccmez2007yet,inci2016cache}, branch predictors~\cite{aciiccmez2007power}, execution ports~\cite{aldaya2019port}, and ring buffers~\cite{paccagnella2021lord}, adversaries can perform sophisticated attacks aiming to leak sensitive information from co-located users in public clouds. Furthermore, cloud providers offer privacy-oriented platforms through Intel SGX and AMD SEV by encrypting the user's application in the memory and executing the applications in an enclave. However, TEEs can also be compromised by malicious tenants and administrators in the cloud through monitoring microarchitectural components~\cite{moghimi2017cachezoom,wang2017leaky,lipp2021platypus,van2019tale}. In both cases, these attacks are relatively well-understood today and mitigation strategies have been enforced to counteract them. For instance, a range of attacks against TEEs was previously possible because of the ability to single-step through enclaves using tools like SGX-Step~\cite{vanbulck2017sgxstep}. However, it is noteworthy that such attacks are no longer feasible, as evidenced by the work of Constable et al.~\cite{constable2023aexnotify}.

In addition to microarchitectural components, specialized architectural registers provide detailed information about device usage. These registers gather data from various sensors, reporting metrics like power/energy consumption, CPU core-specific frequency values, and the number of interrupts. Importantly, these registers can be accessed directly by user-space applications through model-specific registers (MSRs) or the operating system's interface, posing a significant security risk to cloud users. Recently, the Intel Running Average Power Limit (RAPL) interface that reports the device power consumption was compromised to leak secret keys~\cite{lipp2021platypus} and other user activities~\cite{zhang2021red}. Similarly, the \textit{cpufreq} interface was leveraged to monitor user activity through malicious applications~\cite{wang2022hertzbleed,dipta2022df}. These attacks demonstrated that OS-based features are also valuable sources for side-channel attacks.

%Other than microarchitectural components, there are special architectural registers giving more detailed information related to the device usage. 
%These registers collect the information from different types of sensors and report power/energy consumption, CPU core-specific frequency values, and the number of interrupts. On the other hand, these registers can be accessed directly from user space applications through model-specific registers (MSRs) or OS interface, causing a security threat against cloud users. 

In this study, we examine whether containers running in microservices architectures can be identified through frequency reporting sensors available in Intel and AMD CPUs. We focus on the most popular microservices platforms and their underlying systems, such as Amazon Firecracker~\cite{agache2020firecracker} and Google gVisor~\cite{young2019true}, as well as privacy-oriented platforms such as Gramine~\cite{tsai2017graphene} relying on Intel SGX and AMD SEV~\cite{sev2017seves} to assess our attack accuracy for container fingerprinting. (Note that none of these sandboxing mechanisms/TEEs is supposed to protect against dynamic frequency-based side-channel attacks). The acquisition of such knowledge can provide adversaries with the capability to execute more effective, efficient, and stealthy attacks. %Moreover, it can also enable them to detect co-location, a critical prerequisite for performing a wide spectrum of attacks, including Spectre~\cite{10.1145/3399742} and Meltdown~\cite{217478}.

We summarize our key contributions as follows: 
\begin{itemize}
    \item We demonstrate that each Docker image has a unique dynamic frequency fingerprint, and the detection accuracy reaches up to 84.5\% in the native environment.
    
    \item We show that our attack is successful in real-world sandbox (gVisor, Firecracker) and privacy-oriented environments (Intel SGX, AMD SEV) by achieving a detection accuracy of up to 91.4\%.

    \item Our attack can distinguish multiple containers running concurrently with an accuracy higher than 70\%.

    \item We examine the effects of the bootstrap phase of the Firecracker container process, MicroVM creation, and the Docker image pull process on the frequency fingerprints.

    \item Our attack can distinguish different versions of the same image with an accuracy of 81.02\%.

    \item Finally, we propose a noise injection-based defense technique against frequency-based side-channel attacks.
    
    %Finally, we outline a lightweight countermeasure by monitoring system call patterns to detect malicious activities. Our tool detects the attacks with 100\% accuracy and 1.8\% performance overhead.
    
\end{itemize}

\section{Background} \label{sec:background}

We begin by offering the necessary background information to understand the functioning of the Dynamic Voltage and Frequency Scaling (DVFS) present in modern CPUs. Subsequently, we present several sandboxes widely used today alongside two TEE-based solutions that provide enhanced security and privacy for cloud applications.

\subsection{Dynamic Voltage and Frequency Scaling}

Modern processors can operate at different voltage configurations and clock frequencies, commonly known as P-states or Operating Performance Points. Operating a CPU at higher clock frequencies and voltages allows more instructions to be executed during a given time window; however, this comes at the cost of higher power consumption and a significant increase in temperature. From this observation, it becomes evident that the goal is to achieve an optimal balance between CPU utilization and power consumption, using the P-states with the lowest power consumption whenever possible. The use of these P-states is particularly relevant in the context of cloud computing, where server CPUs need not always be running at full speed. Fortunately, today's CPUs integrate multiple hardware interfaces that facilitate automatic switching between various frequency/voltage configurations, dynamically adapting to changing CPU resource demands. Moreover, Turbo Boost technology, present in Intel and AMD CPUs, allows these processors to operate at higher frequency levels than their specified maximum threshold when certain conditions are met, a feature known as dynamic overclocking. Intel devices support this feature through Intel Turbo Boost Technology~\cite{intel_turbo} while AMD enables high operating frequency through Turbo Core technology~\cite{amd_turbo}.

In Linux-based systems, the \textit{cpufreq} subsystem is responsible for adjusting the performance scaling of all CPU cores. This subsystem is structured into three layers: the core, scaling governors, and scaling drivers. Among these layers, the scaling governor is the one in charge of selecting the scaling algorithm that predicts the CPU latency.
Furthermore, it is worth noting that user-space applications can access the current dynamic frequency values of each core, which are updated approximately every 10\,ms and can be read through the cpufreq interface.

\subsection{Sandbox Environments}

%\begin{itemize}
%    \item Short description on the isolation of container in native linux. 
%    \item explain how the Docker images running inside a container environment, should remain private. 
%\end{itemize}
\noindent\textbf{Docker containers~\cite{docker}.} The kernel's resource-sharing model adopted by containers presents significant benefits in performance and efficiency while offering much weaker isolation guarantees. Consequently, it is of utmost importance to protect the host from malicious containers aiming to exploit host OS kernel vulnerabilities. To this end, cloud providers can utilize various Linux kernel functions and modules (e.g., \emph{SELinux}~\cite{Selinux}, \emph{AppArmor}~\cite{Apparmor} or \emph{Seccomp}~\cite{Seccomp}) in order to enforce fine-grained security policies. For achieving isolation among containers on the same host, namespaces and cgroups security mechanisms are typically employed~\cite{7742298}. Namespaces is a key feature in the Linux kernel that enables process and resource isolation by effectively dividing system resources (e.\,g., process IDs or network interfaces) so that each container has its own isolated instance of such resources. Additionally, cgroups governs the allocation of resources (e.\,g., CPU, memory, and disk I/O) to maintain equitable sharing among containers. %Meanwhile, for isolating containers across distinct hosts, it is common to use solutions based on bridge networks, overlay networks and service discovery mechanisms~\cite{254412}.

\noindent\textbf{gVisor~\cite{gvisor}.} Developed by Google, gVisor is an open-source container runtime that adds an extra layer to provide greater security and isolation to containerized applications. Positioned in between the containerized application and the host OS, gVisor serves as a user-space application kernel, intercepting and emulating system calls in order to create individual virtualized environments for each container. This approach mitigates attacks by malicious applications against the underlying host. gVisor goes one step further and extends this protection by isolating itself from the host through well-known Linux isolation capabilities. This multi-layered approach provides defense-in-depth while maintaining performance comparable to virtual machines and efficiency akin to containers. Other key features of gVisor include its minimal privilege requirements, strict system call filtering, and implementation in Go, a well-known memory-safe and type-safe programming language.

\noindent\textbf{Firecracker~\cite{agache2020firecracker}.} Firecracker is an open-source virtualization technology developed by AWS, establishing a minimal and efficient environment for operating lightweight VMs, known as MicroVMs. MicroVMs offer enhanced resource utilization and quicker startup times compared to traditional VMs, while also providing a higher security level compared to Docker containers. Guided by a minimalist design philosophy, Firecracker creates MicroVMs with the minimum essential functions to optimize both performance and security. Executed on the KVM hypervisor, MicroVMs also leverage isolation mechanisms built into the Linux kernel such as Seccomp, cgroups, and Namespaces in order to further improve security within Firecracker MicroVMs. Last but not least, Firecracker integrates a novel virtual machine monitor (VMM) and an extended API for software developers to efficiently manage their MicroVMs.

\subsection{Trusted Execution Environments (TEEs)}

\noindent\textbf{Intel Software Guard Extension (SGX)~\cite{cryptoeprint:2016/086}.}
Intel SGX is a TEE that shields applications by encapsulating them in so-called enclaves. Enclaves are isolated regions of memory protected from the operating system and other software running on the same system. The isolation mechanism ensures that sensitive data and code can be processed within the enclave without exposure to threats or attacks. SGX enables developers to create applications that protect valuable assets like encryption keys, sensitive computations, and authentication processes from being accessed by unauthorized parties. By utilizing hardware-enforced isolation, Intel SGX addresses critical security concerns in various domains, such as cloud computing and mobile phones, securing user data against malicious activities.

%A TEE enables the creation of a secure memory area on the main processor that preserves the confidentiality and integrity of any data and code stored or processed in it.

% \begin{itemize}
% \item 
% \subparagraph{Gramine}
\noindent\textbf{Gramine~\cite{gramine}.} Gramine is a lightweight guest OS designed to execute Linux applications while easing host requirements.
The benefit of Gramine compared to VMs in the cloud is to run applications in an isolated environment with guest customization, process migration, and easily porting the applications to different host OSes.
Moreover, Gramine supports running unmodified applications inside SGX enclaves without manually configuring the application for the SGX environment.
This process is extremely useful for container users without trust in cloud providers since Intel SGX enclaves are protected against adversaries in the host system.
The Gramine Shielded Containers (GSC) tool~\cite{gsc} basically transforms a base Docker image into a graminized Docker image that includes Gramine-specific application configuration and the Gramine Library OS. As a result, Gramine allows cloud providers to run containers in the Intel SGX environment while protecting the application against a malicious host through enhanced security primitives.
% \end{itemize}

%Gramine is a library OS whose goal is to ease the process of running unmodified applications inside Intel SGX~\cite{cryptoeprint:2016/086} while easing host requirements. %The benefit of Gramine compared to VMs in the cloud is to run applications in an isolated environment with guest customization, process migration, and easily porting the applications to different host OSes. %Moreover, Gramine supports running unmodified applications inside Intel SGX enclaves without manually configuring the application for the Intel SGX environment. 
%This process is extremely useful for container users without trust in cloud providers since Intel SGX enclaves are protected against adversaries in the host system. 
%The Gramine Shielded Containers (GSC) tool~\cite{} essentially transforms a base Docker image into a graminized Docker image, including Gramine-specific application configuration and Gramine Library OS. %As a result, Gramine allows cloud providers to run containers in the Intel SGX environment while protecting the application against a malicious host.

\noindent\textbf{AMD Secure Encrypted Virtualization (SEV)~\cite{sev2017seves}.} AMD SEV is a security technology developed by AMD to enhance the protection and isolation of virtual machines in cloud environments. It allows for creating encrypted memory areas called ``encrypted VMs'' or ``encrypted enclaves''. These enclaves isolate the memory and data of each virtual machine from both the hypervisor and other virtual machines, providing a higher level of security against various types of attacks, including those attempting to steal sensitive data from memory. By encrypting virtual machine memory, SEV helps safeguard against threats such as malicious administrators, unauthorized access, and memory-based attacks. This technology is crucial in bolstering the security of cloud computing environments and maintaining the confidentiality of customer data and workloads.

%Google gVisor is the sandbox technology that powers Google Computing Platform’s (GPC) App Engine, Cloud Functions, and CloudML. Google realized the risk of running untrusted applications in the public cloud infrastructure and the inefficiency of sandboxing applications using VMs, and developed a user space kernel to sandbox the untrusted applications. gVisor sandboxes applications by intercepting all the system calls from applications to host kernel and handling them with gVisor's kernel implementation Sentry in user space. It essentially functions as a combination of guest kernel and VMM. gVisor relies on the host operating system and the platform for defense against hardware-based attacks. Host-level mitigations against hardware side channels are still effective with a sandbox. However, gVisor does not add any additional mitigation against the side-channel attacks.
\section{Motivation} \label{sec:motivation}

Our research builds upon the key observation that modern CPUs dynamically adjust their voltage and frequency in response to the workloads they handle, and this information is accessible from user-space. Our hypothesis is that this voltage and frequency information can constitute a unique application's fingerprint. 

In this study, we consider different scenarios in which the victim and the attacker should operate in full isolation within their corresponding domains (container vs user-space or TEE vs kernel-space), i.e., without violating the trust boundaries established in each case. Therefore, regardless of the scenario, adversaries should \emph{not} be able to obtain any information whatsoever about the applications running inside the victim's container. This includes information such as what application is running within the victim container (e.g., a MariaDB instance). Such knowledge could give adversaries valuable information to perform sophisticated attacks more efficiently, effectively, and stealthily. If adversaries can discover that a container implements a MariaDB image, they can design more targeted attacks by taking advantage of known weaknesses and attacks against this type of database. For example, if adversaries know that MariaDB is used, they could try to exploit the remote code execution vulnerability reported in CVE-2021-27928~\cite{cve27928} or the highly successful vulnerability found in CVE-2016-6662~\cite{cve6662}.

%Knowing the version of the application within the container can further help adversaries choose which attack is most effective and efficient – especially in the case where there are known vulnerabilities against specific versions of the application. For example, if adversaries determine that the MariaDB container is utilizing version 10.2, they might attempt to exploit the documented remote code execution vulnerability, CVE-2021-27928~\cite{cve27928}. %Conversely, if the MariaDB container is running version 5.5 or an older release, adversaries may opt to exploit the highly successful vulnerability found in CVE-2016-6662~\cite{cve6662}

%For example, if adversaries know that MariaDB v10.2 is used, they could try to exploit the remote code execution vulnerability reported in CVE-2021-27928~\cite{cve27928}, while if the version of the MariaDB container is 5.5 (or older), a highly successful option for adversaries would be to try to exploit the vulnerability reported in CVE-2016-6662~\cite{cve6662}.
\section{Threat Models} \label{sec:threat_model}
In this study, we consider three threat models: (1) a native environment, (2) a sandbox environment, and (3) a TEE-based environment. In the following, we explain the threat models in further detail.
 
\begin{figure}[t!]
    \centering
    \includegraphics[width=\linewidth]{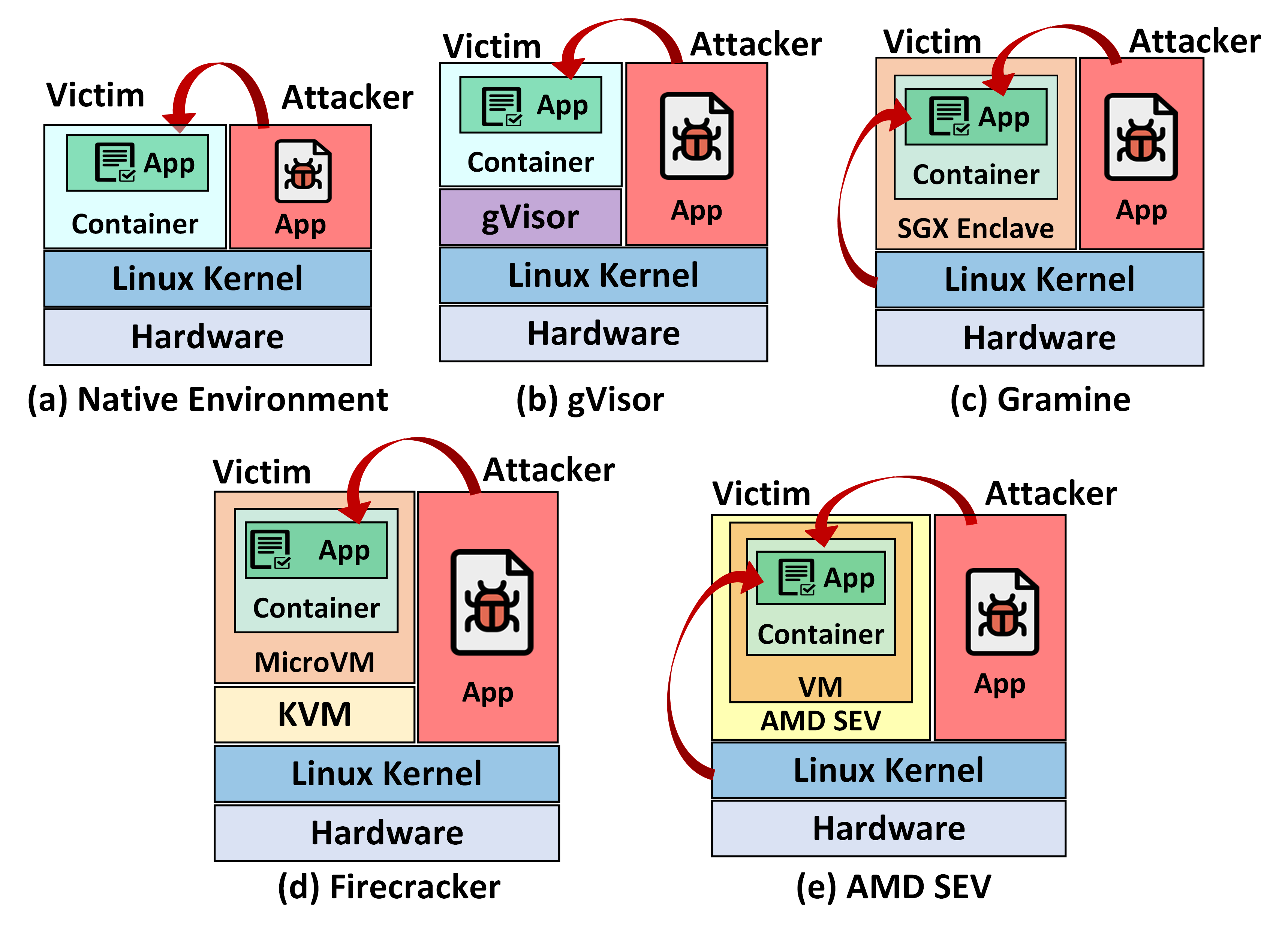}
    \caption{Overview of the threat models of our proposed attack in different execution environments.}
    \label{fig:Threat_model}
\end{figure}

\noindent\textbf{Native Environment.} In the native environment scenario, we assume that the victim is running an application inside a Docker container in the user-space. Concurrently, the attacker, also operating within the user space, seeks to monitor the CPU frequency of all cores in the victim's device through a malicious application or container. (Note that our experiments revealed that, by default, Docker containers expose such CPU frequency information to their owners). In this scenario, the attacker leverages the \textit{cpufreq} interface to gather frequency values and perform fingerprinting of the running containerized applications, as illustrated in \autoref{fig:Threat_model}a.

\noindent\textbf{Sandbox Environment.} In the sandbox environment, we assume that the victim is a tenant running one containerized application in the cloud and the attacker is employed by the cloud provider with user-level access to the host system. The victim runs its containers either using a container-based sandbox (\autoref{fig:Threat_model}b) or inside a MicroVM (\autoref{fig:Threat_model}d). Note that, unlike the previous scenario, the \textit{cpufreq} interface is not accessible from MicroVMs in Firecracker and containers in gVisor). Limiting cloud provider employees to user-level access on its host systems while they are serving customers is a common technique to secure the system~\cite{aws2023nitrosecurity}. However, the cloud provider must be able to gather metrics like CPU utilization, memory allocation, or power consumption to be able to properly bill its customers.

\noindent\textbf{TEE-based Environment.} In this threat scenario, we assume that the victim runs the containers in a more privacy-oriented environment using TEEs like Gramine (see \autoref{fig:Threat_model}c) or AMD SEV (cf. \autoref{fig:Threat_model}e). The victim container is assigned to one core and runs inside an enclave.
Here, we adopt the standard TEE adversary model~\cite{7807249}, which considers
adversaries who have root privileges. The attacker profiles the CPU frequency of the physical core assigned to the victim container. This type of attack can be carried out by the cloud providers themselves, or by malicious tenants who manage to escape the container and gain control of the underlying host~\cite{cve-2017-5123}. In the latter case, after carrying out the attack, the adversary has the same privileges as the cloud provider on that host. 
Finally, we consider two variations for each of the threat models. In the first variant, the attacker is able to capture frequency measurements during the execution of the victim (i.e., from the moment the ``docker run'' command is executed). In the second variant, the attacker can additionally gather measurements while the victim retrieves its container image from a container registry (i.e., while the ``docker pull'' command is being executed). Obviously, the second variant gives the attacker an advantage over the first variant as they have access to more information that can be leveraged for the fingerprinting attack. It is to be noted that, we assume the adversary has a list of official docker images and the fingerprinting attack can profile only the listed docker images. 

\section{Technical Challenges}\label{sec:challenges}
During our work, we overcame three technical challenges (C1, C2, and C3) as described in detail next.

%related to data collection
\noindent\textbf{\textit{C1: Collecting container frequency signatures.}}\hfill\\
%\eduard{Here we should explain the challenges we encountered when reading the registers to obtain the frequency signature. We can talk about the noise introduced by other applications running in the server (we can for example say that each core has its own frequency measurements). Another challenge could be associated to whether the adversary sees the docker pull and docker run or can only see one of them. Following this, we could also stress that we want to study the feasibility of these attacks when the adversary is "weak" and can only monitor the container for a short period of time.}
Adversaries aim to devise an efficient and effective approach for accurately acquiring the frequency signature of a (victim) container. Achieving this goal poses several challenges, as follows. One such challenge is how to disentangle the frequency pattern of the underlying execution environment from that of the container image. This complexity is particularly pronounced when Docker containers operate within MicroVMs, such as those generated by Firecracker. Moreover, while collecting frequency measurements, the container operates within a non-isolated core, introducing noise into the measurements due to concurrent tasks sharing the same core. Another potential hurdle for adversaries relates to the specific type of frequency measurements they have access to. In certain scenarios, unauthorized access to the server might have been obtained by adversaries before the container is executed, enabling them to monitor the frequency patterns produced as a result of the ``docker pull'' and ``docker run'' commands. In other cases, however, adversaries could compromise the server after the image has been pulled and stored on the server, meaning they can only see the frequency signature while a container is running.

\noindent\textbf{\textit{C2: Distinguishing container frequency signatures.}}\hfill\\
The second challenge lies in accurately identifying the application running inside a container solely from its frequency signature, taking into account that it is common for similarities between containers to exist. It is worth noting that Docker images are made up of layers--base, intermediate, and top layers. Many of these layers are shared across containers to optimize storage and transfer efficiency. For instance, base image layers include operating system files and core software components that can be directly used in many containers. Another example is that multiple containers may share software dependencies such as libraries, packages, and frameworks, leading to the same image layers. Complicating matters, software developers frequently rely on Docker images sourced in public repositories as foundations to build their own applications, thereby exacerbating the layer-sharing problem and resulting in more similar frequency fingerprints between containers.

\noindent\textbf{\textit{C3: Exploiting the leakage arising from the frequency signatures to compromise privacy.}}\hfill\\
Once adversaries have been able to collect the containers' frequency signatures accurately and use the collected measurements to distinguish them from one another, they have to think of ways to carry out their attacks in the real world. In a real cloud computing environment with many tenants, each concurrently running numerous containers, adversaries must possess the capability to monitor all cores of the server they compromised. Swiftly identifying the creation of a sandbox environment designated for application execution becomes crucial. Additionally, adversaries consistently prioritize launching attacks that yield substantial impact while maintaining cost efficiency. With this in mind, we are interested in investigating the feasibility of the proposed fingerprinting attacks by ``weak'' adversaries who are limited to collecting frequency measurements only for a short amount of time. %\thore{Rate-limiting the attacker doesn't match with threat models 2 and 3 which assume the CSP as the attacker.}

%\eduard{Here one of the challenges would be to identify the core in which the container is being executed. Another challenge could be how to perform such fingerprinting attacks in different types of execution environments. As we need to train a diff ML model for each environment, we can say that the adversary could easily know which environment cloud providers use as this is information is publicly available online }
\section{Methodology}\label{sec:methodology}
%\eduard{We can try to write the steps we followed to analyse the feasibility of fingerprinting attacks in distinct types of execution environment. A few examples:}

%\noindent \eduard{1- Crawling Docker images from public repositories} \\
%\noindent \eduard{2- Running the container and identifying the core where it is executed} \\
%\noindent \eduard{3- Monitoring the frequency signature of containers} \\
%\noindent \eduard{4- Feature extraction/selection/reduction} \\
%\noindent \eduard{5- Preparing the ML model} \\
%\noindent \eduard{6- ...} \\

% \berk{online and offline steps described here}
% The entire methodology can be categorized into different steps to analyze the feasibility of our attacks. As mentioned in the threat model, the attack consists of offline and online phases. Most of the required steps are performed in the offline phase, while the online phase can be considered as a testing stage of our experiment. All of these steps are summarized below: \\

In all the proposed threat models, the attack consists of two distinct phases: an \emph{offline} phase, which occurs before the victim container is executed, and an \emph{online} phase, which takes place while the victim's container is running. In the offline phase, the attacker collects CPU frequency data through the \texttt{cpufreq} interface while running different Docker images in a container separately. To that end, the attacker can either use container images they own or rely on widely used container images located in public repositories such as Docker Hub. The collected CPU frequency data is then used to train a deep learning model, creating a pre-trained model for the online phase of real-time fingerprinting attacks. During the online phase, the CPU frequency readings of the victim's application are recorded by a malicious app and are forwarded to the attacker's device through a communication network. The attacker utilizes the pre-trained model to predict the running Docker image based on the frequency fingerprint. %In addition, we target one CPU core while running the Docker images in the victim's device. 
Note that most of the required steps are performed in the offline phase, while the online phase can be considered as a testing stage of our experiments. In the following, we describe the steps performed during both phases in further detail.

%We assume that only one container runs at a time in the victim's device 
%\eduard{(Don't we have the means to collect frequency information for each of the cores? If that is the case, I don't understand why we need to assume this (of course, considering that there is 1 container per core))} while collecting the CPU frequency data. 

%\noindent \eduard{(I made some changes here but I'll come back once I've read the rest of the sections in the paper)}

%\smallskip
\noindent\textbf{1) Retrieve container images.} 
% To demonstrate that our attack is possible, we initially focus on an environment featuring Docker containers executing Docker images sourced in Docker Hub, the most popular repository of Docker images that exists today.
% We start by developing a custom web scraper to gather the information needed to run Docker Hub images leveraging the Docker Hub API~\cite{DockerHubAPI}. We configure the crawler to retrieve all official Docker images hosted on Docker Hub, which at the time we performed the crawling, numbered 178. Subsequently, we identified that 23 of these images were either deprecated or obsolete, leading us to exclude them from our dataset. Unless otherwise specified, any experiment discussed in the rest of this paper considers the remaining 155 Docker images. \\
During the initial step, which takes place in the offline phase, the attacker obtains the container images they want to use in their dataset. This can involve retrieving container images from a public container image repository or constructing the container images from source code repositories. For our experiments, we developed a custom web scraper to gather public Docker images from Docker Hub through the Docker Hub API~\cite{DockerHubAPI}. Our crawler was configured to retrieve all official Docker images on Docker Hub, which resulted in a total of 178 container images. Out of the initial set of container images, we excluded 23 images that were either deprecated or obsolete. This process resulted in a dataset containing 126 Docker images, which we used for our subsequent experiments and analysis.

%We consider the official Docker images for this experiment and pull them from Docker Hub. Although 178 official Docker images are available in the repository, we could not run all of them. Of these 178 official images, 23 Docker images do not run as they are either deprecated or obsolete. Hence, the experiment is carried out with 155 official Docker images for the Native Linux Environment. 

%\noindent\textbf{2) Running Docker containers:} The second step involves individually running each of the containers with their arguments. In our study, we adopt the minimum set of arguments essential for the proper working of the Docker container. We make this choice because we view it as a worst-case scenario for potential adversaries. It is plausible that providing a more extensive set of arguments to the container would result in a more distinctive frequency signature, ultimately making the container's fingerprint more unique. \\

%We speculate that this is a worst-case scenario for the adversary, since passing a richer set of arguments to the container is likely to produce a more unique frequency signature, making the container's fingerprint more evident.

%In this study, we used the minimum set of arguments needed for the Docker container to run. If the containerized image can be detected with these minimal arguments, it will be a success indeed, considering a weak adversary. It is to be noted that the arguments for running each Docker container remain consistent over different environments that have been explored in this study.

%\newpage
\noindent\textbf{2) Converting container images into different runtimes.}
In the next step, the attacker prepares the Docker container images for the specific target execution environment, which can vary depending on the sandbox on which the attack will be performed. For the container environment, this step is rather straightforward if the retrieved container images follow the specifications of the open container initiative. When VMs are used to isolate container workloads, the images first have to be converted to a virtual machine format. This is usually done by either converting the container image into a bootable root file system by flattening the image layer into a single file system, adding a kernel, and configuring the initialization of the container workload~\cite{salmen2021containervmconversion} or the container image is moved into a prepared VM that has a container runtime installed which will execute the container inside the VM~\cite{firecracker2023buildfcvm,kata2023containervmconversion}. If the execution environment is a process-based TEE like SGX, the container image must be converted into an enclave for that TEE. Execution environments that provide a TEE for VMs like AMD SEV or Intel TDX require the container-to-VM conversion, followed by a conversion from a generic VM image to a TEE enclave. The attacker performs these steps during the offline phase of the attack.

% Consider, for example, Firecracker and AMD SEV. To run a container workload like, e.\,g., an nginx Docker container with Firecracker, the layers of the container image have to be flattened into a single file system \eduard{(@Thore: Can you add a reference here?)}. Additionally, a kernel has to be added which will be used to boot the MicroVM, mount the container file system and start the nginx task. With AMD SEV, the whole MicroVM image must also be encrypted and--optionally--cryptographically signed to allow for remote attestation of the fact that the container workload was loaded and initialized as expected. The attacker performs this step during the offline phase of the attack.

We use a Firecracker-specific version of containerd to run Docker images inside microVMs via a container runtime inside the VM, prepare the images for SGX with Gramine, rely on the Kata container runtime when working with AMD SEV, and integrate the \texttt{runsc} runtime with containerd for gVisor. More information about the several setups and conversion steps are given in the respective sections later in the paper.

%\thore{anything special for gVisor?}\debopriya{addressed}
\smallskip
\noindent\textbf{3) Collecting CPU frequency measurements.}
%The third step involves monitoring the frequency signature of each Docker for several iterations. The Linux kernel provides the current CPU frequency of each virtual core through the \texttt{scaling\_cur\_freq} attribute. 
To collect CPU frequency signatures for individual Docker containers, the adversary monitors the CPU frequency (\texttt{scaling\_cur\_freq} attribute) during the runtime of each container. A generalized data collection algorithm\footnote{The dataset and the code will be made available on GitHub: {\url{https://github.com/Diptakuet/Dynamic-Frequency-Based-Fingerprinting-Attacks-against-Modern-Sandbox-Environments}}} is presented in \autoref{alg:data_collection} that works for all environments. The algorithm reads the CPU frequency ($f$) in parallel while running a specific container. The sampling rate ($T_i$) and the number of samples ($N_s$) in each measurement are predefined for each environment. This step is performed during the offline phase to gather training data for the ML model and also during the online phase to retrieve measurements from the victim container. In general, we collected 4000 samples with a sampling rate of 10\,ms (unless otherwise specified), which takes 40 seconds to collect one measurement. %We collected 100 measurements for each Docker image in order to create our dataset during the offline phase.
If the number of images is N and the number of measurements per image is M, the overall data collection time during the offline phase takes around $(N \times M \times 40)$ sec. 
Further details on chosen Docker arguments are given in Section 9.

%For our experiments, we intentionally limit the monitoring time of the container's execution to a few seconds and use the minimum set of Docker arguments required for the proper operation of the Docker container. We have made these choices to demonstrate the attack's effectiveness in the worst-case scenario for potential adversaries. Note that, providing a broader set of arguments to the container and considering a longer monitoring time will most likely only make the container's fingerprint more evident and thus that the attack easier to perform. 

\begin{algorithm}[t!]
\scriptsize
\caption{Data Collection Algorithm}
\label{alg:data_collection}
\tcp{$T_i$ is the interval between each reading}
\tcp{$N_s$ is the number of samples}
\tcp{$N_C$ is the number of containers}
\tcp{$N_M$ is the number of measurements per container}
\tcp{$Container\_name$ is the name of the container}
\tcp{$f$ is the CPU frequency}
\KwInput{$T_i,N_s,N_C,N_M,Container\_name$}
\KwOutput{$f$}
%begin{algorithmic}
\For{$i \gets 1$ to $N_C$}{
    \For{$j \gets 1$ to $N_M$}{
          Run $Container\_name[i]$ \; 
          \For{$k \gets 1$ to $N_s$}{
               $f[i,j,k] \gets$ Read $scaling\_cur\_freq$ \;
               sleep $T_i$ ;
            }
          kill $Container\_name[i]$ \;
          sleep $5s$ \;
     }     
}
%\end{algorithmic}
\end{algorithm}

\smallskip
\noindent\textbf{4) ML model deployment.}
In the fourth step, we train a Convolution Neural Network (CNN) model to establish a pre-trained model based on the CPU frequency data collected in the previous step. 
The model architecture consists of multiple convolutional layers, max-pool layers, dense layers, and dropout layers. The \textit{ReLU} activation function is used for all convolutional and dense layers except for the last one. The last dense layer incorporates the \texttt{softmax} activation function. The dropout rate in the dropout layers is set to 0.5. We also utilized a fixed kernel size of $3 \times 1$ for all the convolutional layers. The only parameters that are changed during each specific execution environment are the number of convolutional layers and the number of neurons in the dense layers. %\thore{From here, the paragraph is partially redundant with the previous steps and also already mentions content of the next step. Rather include some specifics about the model we trained (hyperparameters like the number of layers, the number of neurons, ...)}\debopriya{addressed} 
Note that this ML model is only able to accurately classify samples from any of the containers included in the dataset. The time for building a pre-trained model in the offline phase depends on the computational power of the adversary.  We used NVDIA GeForce 3090 GPU to create our pre-trained model. The time to train a model takes around 1.5 minutes in our experimental setup.

%To evaluate its performance, the pre-trained CNN model is tested with CPU frequency data monitored during the online phase. 
\smallskip
\noindent\textbf{5) Conducting fingerprinting attacks in the wild.}
Once the adversary has developed an ML model capable of accurately identifying images in its training dataset solely based on their frequency signatures, the next step is to consider the practical execution of the attack (i.\,e., the online phase). To achieve this, the adversary must possess knowledge about the sandbox employed by the victim application and the specific core it is utilizing. For the former, the adversary can leverage the fact that information regarding execution environments and the micro-architecture types for different instances used by major cloud providers is usually well-documented~\cite{aws_instance, aws_amd, amd_sev, alibaba_ecs, amd_google_cloud}.
For the second case, it becomes crucial to determine whether the sandbox frequently employed for running cloud-based applications also possesses a unique fingerprint that can be used by adversaries to detect the presence of a new container. Afterward, the attacker collects a CPU fingerprint from the victim's device and feeds the fingerprint into the pre-trained model to classify the running container. %\thore{Do we consider the inference/classification as a part of the online phase or do we only collect frequency data and later (during another offline phase) classify the data we collected earlier?} \debopriya{Do we consider the inference/classification as a part of the online phase? YES}

%\noindent\textbf{4) Identifying the core:} In the cloud environment, usually each user is assigned a specific core of a server. If one of the users runs the container in a specific core, the adversary first needs to identify the core where it is executed. ..

%\noindent\textbf{7) Setting up Execution Environment:}
%\eduard{We could have this as step 7}

\section{Attack in the Docker Environment}\label{sec:docker}
%\subsection{Identifying the core}
%\eduard{The current title may not be the best, but I think it would be good to present our work/results following the logical attack sequence}

%Every docker image consists of multiple layers, which are basically read-only files. When a docker image is run inside a container, a top layer is created on top of the immutable layers. This top layer is writable by the users. Every process running inside a container is completely isolated from the host OS, ensured by the native Linux kernel. However, the main concept of this research initiated with an assumption that the building layers of the docker images running inside a container will create unique fingerprints at the microarchitecture level. Therefore, the isolation of docker containers could still be vulnerable to microarchitecture-based side-channel attacks. 

Our work is built upon the hypothesis that \textit{a series of CPU frequency samples collected while a container is in operation can develop into a reliable container fingerprint}.
To confirm this hypothesis, we choose to employ DVFS in order to quantify the fluctuations in CPU frequencies during the execution of Docker containers. In this section, we perform our attack on a native Linux environment. We built a dataset of 126 Docker images and trained a CNN model to evaluate the performance of our attack. The attack execution steps are explained in detail in light of the attacker.

%\eduard{(Add 1 or 2 sentences to explain what comes next in this section)}

%\subsection{Executing the attack}

\noindent\textbf{Experiment Setup.}
The experiments for the native Linux environment are performed on the Intel Comet Lake microarchitecture with a CPU model of Intel Core i7-10610U CPU @ 1.80GHz. The installed version of containers and docker-init are v1.6.6 and 0.19.0, respectively. The OS is Ubuntu 20.04 LTS with a Linux kernel version of 5.11.0-46-generic. Intel Turbo Boost is enabled.

\begin{figure}[t!]
    \centering
    \includegraphics[width=\linewidth]{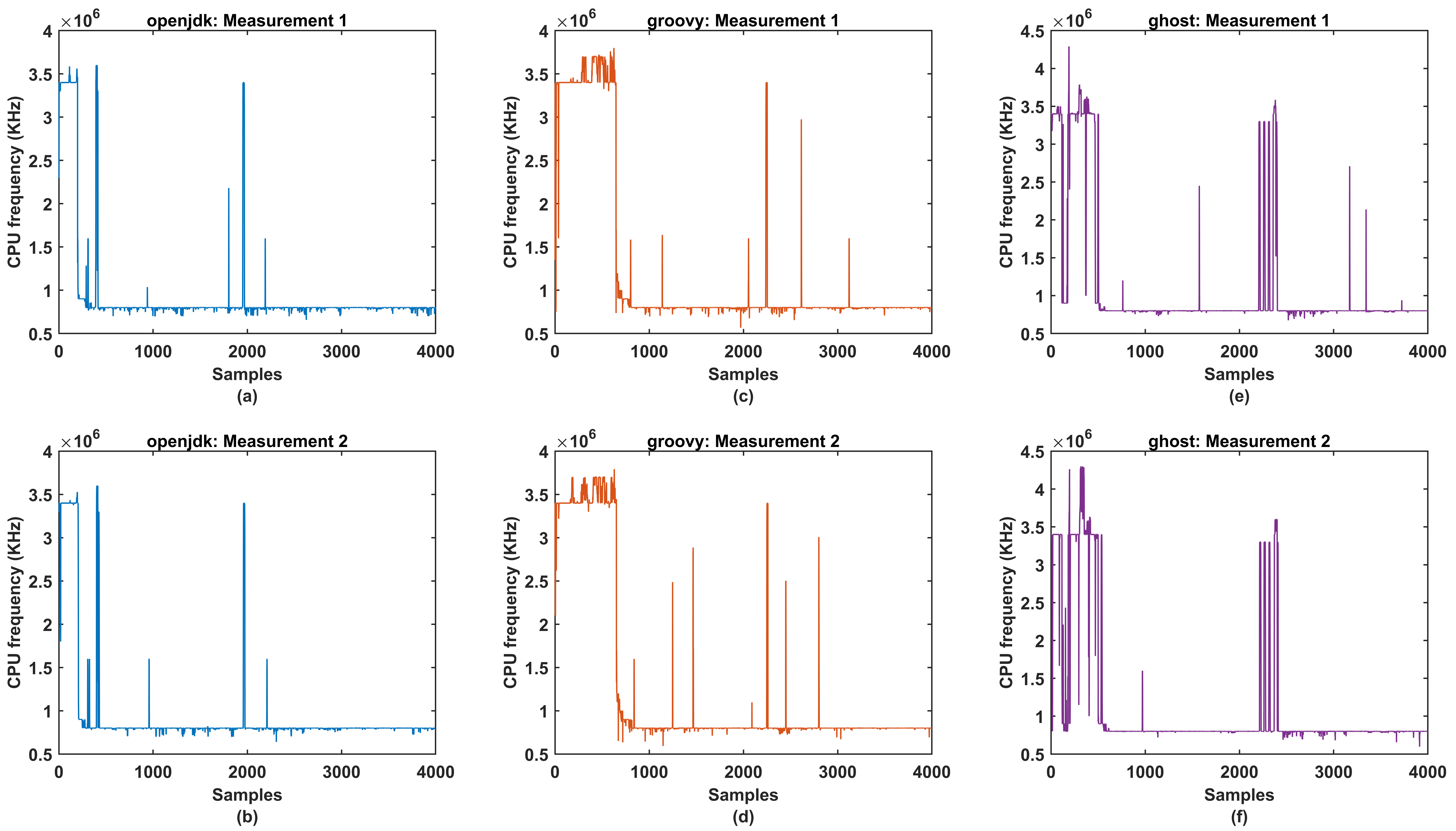}
    \caption{Frequency signatures of containers running in the Native Linux environment are given. \texttt{openjdk} (a, b), \texttt{groovy} (c, d), and \texttt{ghost} (e, f) containers have distinct fingerprints.} %Here, we can observe that the individual frequency fingerprints of these three images are mutually distinct. However, a unique pattern exists for multiple measurements of the same container.}
    \label{fig:native}
    \vspace{-5mm}
\end{figure}

%\subsection{Case Study 1: Native Linux}
\noindent\textbf{Data Collection.}
To evaluate the feasibility of the attack, we collected 100 measurements $(N_M)$ for each container, resulting in a dataset containing a total of 12600 measurements. Each measurement consists of $N_s=4000$ samples, collected at the $T_i=10$\,ms resolution. Hence, one fingerprint is captured in 40 seconds. In \autoref{fig:native}, we visualize six frequency fingerprints collected from three different containers. Note that, while a consistent pattern is observed across multiple measurements of the same container, the frequency signature varies among three individual containers. We also observed that while some containers result in higher operating frequency for a longer time (as in the case of \textit{groovy}), some containers possess less activity (such as \textit{openjdk}).

%We collected 100 measurements $(N_M)$ per container. Thus, the dataset contains $155 \times 100= 15500$ measurements in total. Each measurement consists of $N_s=4000$ samples that are collected with a resolution of $T_i=10$ ms using the \texttt{scaling\_cur\_freq} attribute. In \autoref{fig:native}, we have illustrated four sample measurements from two different containers. We can observe that a common pattern exists for multiple measurements of the same container. However, the frequency signature is different for two different containers.

%\noindent\textbf{Result:}
%\eduard{It would be nice to add more details about the ML part. Which parameters (type of layers) did you use? Cross-validation? Did you apply any feature reduction techniques? }

\noindent\textbf{Attack.}
The collected CPU frequency measurements are one-dimensional. The CNN model consists of three convolution, two max-pooling, and three dense layers. Additionally, two dropout layers are added in between the convolution and max-pooling layers to avoid overfitting. %The kernel size for all the convolutional layers is set to $3\times1$. The \textit{ReLU} activation function is utilized for each convolution and dense layer except for the last one. The last dense layer utilizes \textit{softmax} as the activation function. 
For the optimizer, we utilize categorical cross-entropy since we have more than one class. Out of 100 measurements per container, we randomly chose 60 measurements for training, 20 measurements for validation, and 20 measurements for the online phase (test data). %Thus, $80 \times 155 = 12400$ measurements are used to create a pre-trained model, which is tested on the online phase in real-time with $20\times 155 = 3100$ test measurements in total.
The obtained test accuracy with the CNN model is 84.5\%. This result shows that \textit{our hypothesis is valid in the scenario of identifying the applications running inside Docker containers in a native Linux environment with relatively little noise.}

%In other words, we assume that the container remains consistently assigned to a single core and does not migrate to other cores during its execution.

Our attack relies on two assumptions for its success. Firstly, we assume an exclusive core usage scenario, where each core hosts only one active container at any given time. Secondly, we presume a stable core assignment throughout a container's operational lifetime. 
Due to the limited number of samples required for the attack, we work under the premise that the container stays on a core for a short period, typically a few seconds, which we consider a reasonable premise. Considering these two assumptions, the first piece of information that the adversary needs to know is identifying the core in which the containers are running. One solution for such a problem is to monitor the CPU frequency of all the cores and analyze the signatures with a separate ML model that closely resembles the fingerprints of the listed containers. The confidence rate of the prediction for the fingerprints of each monitored core can be utilized to address this issue. If we consider a realistic execution environment for the containers, identifying the core is comparatively easier than the native Linux environment, as in most cases, the execution environments create a unique signature. 

%We have two assumptions to implement a successful attack against a victim: 1) only one container will run at a time in each core (i.\,e., we exclude the scenario where multiple containers are running in the same core) and 2) the container will always run on the same core, i.\,e., the workload will not be executed on more than one core throughout its live span. 

%\eduard{(Here we can explain the part related to the identification of the core)}
\section{Attacks in Modern Sandbox Environments}  \label{sec:case_study}

In this section, we go beyond the previous native scenario and demonstrate the applicability of our attack in more practical and real-world execution environments like gVisor and Firecracker, which are used by millions of users in production environments today. (Note that in the sandboxes based on creating user-level microVMs, we employ Docker containers inside them).

\subsection{gVisor}

As a first step, we set ourselves the objective of evaluating our attack in gVisor, a popular sandbox that allows the creation of containers with much better security guarantees than those of Docker.

\noindent\textbf{Experiment Setup.}
The experiment for gVisor is conducted on a server equipped with Intel Xeon Gold 5218 CPU @ 2.30GHz. The device has 32 cores with 16 cores per socket. The host OS is Ubuntu 20.04.6 LTS with a kernel version of 5.15.0-78-generic. The version of the installed containerd tool is v1.6.21. Note that this hardware is a typical high-end server that is commonly used in cloud environments.

\noindent\textbf{Data Collection.}
%\eduard{\st{gVisor provides an additional isolation layer between the host OS and the application. It allows the user to run sandboxed containers through \texttt{runsc}, which is an Open Container Initiative (OCI) runtime. The containerd in conjunction with the \texttt{runsc} runtime, is used to manage the sandboxed containers.}} \eduard{(To me this fits better in the background section (and I think we have already explained this there))} 
In order to set up the gVisor execution environment, two crucial components are required. The first one is the \texttt{runsc} runtime through which gVisor manages the runtime interface and runs containers in isolated sandboxes. The other required component, named \texttt{containerd-shim-runsc-v1}, is a shim plugin that works as an intermediary between \texttt{runsc} and containerd. Specifically, this plugin is responsible for managing the communication interface between containerd and \texttt{runsc}. Once the execution environment for gVisor is ready, any containerd CRI tool can be used to run containers in the gVisor environment. In our experiment, the \texttt{ctr} command line tool~\cite{containerd} carries out the tasks associated with pulling and running of the sandboxed containers with the integration of \texttt{runsc} runtime. Namely, it is equivalent to the functionality provided by \texttt{containerd} when Docker containers are used.

%\debopriya{addressed}\berk{Can we comment on this a bit further?}

\noindent\textbf{Results.} %\eduard{(Be careful here -- a Virtual Machine (VM) can be considered a type of sandbox as well. Better to use "container technologies")}
The CPU frequency data is collected for 126 official Docker images selected from Docker Hub~\cite{DockerHubAPI}. %We could not run the rest of the official images in the gVisor environment due to some specific issues. For example, some of the Docker containers show unknown status while running them in the gVisor environment. 
We collected 100 measurements for each Docker image consisting of 4000 samples. The same arguments from the native Linux environment scenario are used. A new 1D CNN model is built for gVisor, which is trained with a randomly selected dataset (60\% training, 20\% validation) in the offline phase. Unlike the previous CNN model for the native environment, an additional convolutional layer is incorporated in the CNN model. Thus, the CNN model includes four convolutional, two max-pool, two dropouts, and three dense layers. The kernel size remains the same as $3 \times 1$ for each convolutional layer. In the online phase, the pre-trained model is tested with 20\% of the data that is not included in the training phase. The obtained test accuracy of the CNN model is 71.2\%. The sandboxed containers possess an extra layer of isolation and are protected against many microarchitectural attacks. Still, our results indicate that the container technology offered by Google cannot protect the identity of the running containerized application against dynamic frequency-based side-channel attacks.

\subsection{Gramine} 
%\eduard{\st{Gramine framework leverages Intel SGX to provide hardware-based isolation for the user to run secured computations. This framework allows the user to create and manage enclaves more seamlessly by extending the capabilities of SGX}}. \eduard{(Let's try to keep all the background info in the background section :))} 
We also explore whether it is possible to fingerprint the Docker images running inside Gramine. Much like Docker and gVisor, Gramine operates as a container-based sandbox; however, it distinguishes itself by running containers backed by Intel SGX, thereby offering enhanced security assurances. It is to be noted that the Docker images running inside the enclave are built from the public images available in Docker Hub. Also, note that the extracted fingerprints do not directly correspond to the characteristics of the public Docker image; instead, they refer to the signatures of the newly built signed-graminized images. 

%\eduard{(As much as possible, let's try to use the same structure that we used previously)}

\noindent\textbf{Experiment Setup.} We perform this experiment on an Intel Coffee Lake microarchitecture compatible with Intel SGX. The CPU configuration of the mobile server is Intel Core i9-9980HK CPU @ 2.40GHz. The installed OS in this device is Ubuntu 22.04.3 LTS with a kernel version v5.15.0-86-generic.

%Intel(R) Core(TM) i9-9980HK CPU @ 2.40GHz
% Ubuntu 22.04.3 LTS
% 5.15.0-86-generic
\noindent\textbf{Data Collection.} The first step before using the Gramine framework is to enable Intel SGX from the BIOS setup. The necessary SGX driver requires to be installed as well. Once the device is ready to use Intel SGX, the second step is to install and configure the Gramine framework. The Docker images available in the public repository cannot be used directly to run inside a protected enclave. Gramine framework provides a feature that can convert any pulled image from the public repository into an unsigned-graminized image. (Optionally, this unsigned-graminized image can be converted again into a signed-graminized image).
In this experiment, we convert all the public Docker images into their signed-graminized version. Afterward, these images are run inside an enclave and can be protected from any untrusted part of the system, including the OS and the hypervisor. We considered 50 images for this experiment which were converted into signed-graminized images as the largest enclave size supported in our target device is 256\,MB. Once the graminization was complete, we ran the signed images inside an enclave and collected the CPU frequency data in parallel. This experiment proves the efficacy of the attack to distinguish signed graminized Docker images which are built from the official Docker images. 

\noindent\textbf{Results.} We collected 100 measurements for each image and trained a CNN model to make a prediction. The pre-trained model can predict the containerized images with an accuracy of 91.38\%. We observed higher accuracy for detecting the graminized images compared with the detection of official Docker images in other execution environments. The graminized Docker image incorporates additional layers on top of the existing layers of the official Docker image from which it is built. This feature might provide some unique activity during the runtime of the individual graminized image to affect their frequency signature distinctively.
%We observed that the fingerprints collected from Gramine framework are more noisy than the native Linux and gVisor environment due to the additional security features of enclaves. Moreover, the setup phase of containers takes more time than in previous scenarios, leading to more noisy measurements. 
%\noindent \textbf{Attack.} \\

\subsection{Firecracker VMM}

In this section, we perform our fingerprinting attack in a Firecracker environment in which MicroVMs are utilized to run containers in a sandbox environment.

\noindent\textbf{Experiment Setup.} 
To ensure a fair comparison, we maintain an identical experimental setup for Firecracker as the one employed for gVisor. In addition to the previous features, the device is compatible with KVM kernel. The installed version of the Firecracker framework is v1.1.0.

\noindent\textbf{Data Collection.}
Firecracker VMM allows clients to create multiple MicroVMs in the cloud provider's server. Inside each MicroVM, the client can run any containerized Docker image. In this experiment, we consider that the client is assigned to a specific core of the server to create a MicroVM and run the containers inside it. We also assume that the MicroVM is always run in the same physical core.%\eduard{\st{The malicious app is assumed to be hidden in the server, which can continuously track the CPU frequency of the system.}} \eduard{(This is also the case for the other runtime environments, so we just need to mention it once in the threat model)} 

\texttt{firecracker-containerd} is the containerd binary used with Firecracker through which the containers can be managed inside the MicroVMs. This process is also equivalent to the Docker daemon established in the native Linux environment. The initial step before running a containerized application inside a MicroVM is to run the \texttt{firecracker-containerd} daemon with a pre-built configuration file. The configuration file defines the initialization of the containerd runtime. For this experiment, the considered configuration script to allow certain settings is summarized in Appendix~\ref{appendix:4}.

\begin{comment}
\begin{itemize}[leftmargin=*,noitemsep]
  \item \textbf{Disable plugins:} The plugins are simply extensions that are supposed to enhance specific \texttt{firecracker-containerd} capabilities. The \texttt{io.containerd.grpc.v1.cri} plugin is disabled through the configuration file, i.\,e., this certain container runtime feature is excluded from the setup.

  \item \textbf{Storage and State Management:} The configuration file defines the root directory for firecracker-containerd's storage. The state information of the firecracker-containerd is also stored in a separate directory specified by the configuration file.

  \item \textbf{Configuration of gRPC setting:} gRPC is a high-level communication protocol that ensures communications between services. 
  One of the best capabilities of gRPC is its multiplexing capability that works across different languages and platforms. In the configuration file, the address of the Unix domain socket is specified to ensure seamless communication for gRPC.

  \item \textbf{Configuration of snapshot plugin:} The \texttt{io.containerd.snapshotter.v1.devmapper} plugin is configured to utilize the Devmapper technology. This plugin specifically helps to manage the container images effectively. To configure this setting, a storage space of 10\,GB is allocated for the efficient management of the container images. 

  \item \textbf{Debug capability:} The debug capability is added through the configuration file by setting the log level which gives an opportunity for troubleshooting.
  
\end{itemize}
\end{comment}

After creating the script, the second step is to configure the device mapper thinpool, which is used by the firecracker-containerd as a storage driver. The device mapper thinpool has two components--data and metadata. A configuration file is created that generates the ``data'' file with a size of 100\,GB and the ``metadata'' file with a size of 2\,GB. These two components allow an optimized and efficient infrastructure to manage the dynamic data and operational metadata of the MicroVM.

The last step is to set up the \texttt{aws.firecracker} runtime, which is required to run Firecracker MicroVM. This configuration file mainly includes the path to the \texttt{firecracker} executable binary, kernel image, and the root driver image. The kernel image facilitates the boot-up phase of the MicroVM, and the root driver creates its file system. The configuration file sets up the necessary kernel arguments, which are passed into the MicroVM. The network interface with the MicroVM is also established through the configuration file setup.

Once we run firecracker-containerd, the containerd runtime gets initialized and booted in within a very short period ($\approx 50-80$\,ms in our test setup). Later, we utilize the \texttt{firecracker-ctr} tool to manage the containers inside the MicroVM, which is equivalent to the \texttt{docker} command in the native Linux. The \texttt{firecracker-ctr} tool allows users to start a MicroVM and manage the containers/images inside. In other words, the firecracker-ctr tool allows running containerized applications encapsulated by the MicroVM.

Before starting the data collection, we run firecracker-containerd in a separate terminal. In a new terminal, \texttt{firecracker-ctr} tool is used to pull and run all the official Docker images. We adopt a similar data collection algorithm as given in \autoref{alg:data_collection} and collect CPU frequency data during the runtime of the containerized applications. We collected the fingerprint dataset for 126 official Docker images in this manner. %For each Docker image, 100 measurements are collected as before. We could not collect data for the rest of the official images as we could not run them successfully as a containerized application inside the MicroVM using the \texttt{firecracker-ctr} tool. 

\noindent\textbf{Results.} For Firecracker, a CNN model is trained (60\%) and validated (20\%) with the collected dataset in the offline phase. The architecture of the CNN model remains the same with the gVisor scenario. The test accuracy of the pre-trained model in the online phase based on the test dataset (20\%) is 73.04\%, which is similar to gVisor. If we consider the top 3 and top 5 guesses of the prediction model, the accuracy becomes 81.2\% and 86.4\%, respectively. Although Firecracker is designed to provide two layers of isolation through container and MicroVM, our results show that Firecracker VMM is still vulnerable to dynamic frequency side-channel attacks.

\subsection{AMD SEV}
In this scenario, we intend to evaluate our attack in an AMD SEV execution environment that is regarded as one of the most secure execution environments in the cloud. %As mentioned in \autoref{sec:background}, AMD SEV provides hardware-based memory encryption to isolate VMs. 
%If the container runs inside AMD SEV-protected isolated VMs, we would like to explore the feasibility of our attack to fingerprint the running container in this section.

\noindent\textbf{Experiment setup.}
We ran our SEV experiments on an AMD EPYC 7232P (Zen 2) CPU @ 3.1GHz processor that features eight cores and 16 threads. 
The CPU is capable of SEV and SEV-ES.
The host OS is an Ubuntu 22.04.3 LTS with a kernel version of 5.15.0-78-generic.
To activate SEV, we run the host kernel with parameters \texttt{kvm\_amd.sev=1} and \texttt{kvm\_amd.sev\_es=1}.
We use containerd version 1.7.3 and the Kata container runtime version 3.2.0-rc0\footnote{Earlier (stable) versions only supported SEV-SNP which is not supported by our CPU.} to run container workloads in SEV-secured virtual machines.
The Kata runtime uses QEMU version 7.2.0 (kata-static).
QEMU is configured to run the VMs with eight host vCPUs which are made available through KVM. The kernel version inside the VMs is 5.19.2-112-sev.

\noindent\textbf{Data Collection.}
The SEV feature of the AMD requires to be enabled from the BIOS setting in the first stage. Later, we configure QEMU to create and run a SEV-protected VM. In order to run the container inside the VM, the Kata container runtime is utilized. In this experiment, we use the \texttt{nerdctl} command line tool to manage the runtime and the features of containerd. We collected CPU frequency data from 107 official Docker images in the Docker Hub~\cite{DockerHubAPI}. We faced some difficulties while running the rest of the official Docker images, leading to a low number of images compared to previous scenarios. We collected 4000 samples with 10\,ms resolution for each measurement and created a data set that contains 100 measurements of each Docker image. 
% BIOS settings and enable SEV support.
% QEMU is configured to run VMs
% Create a new virtual machine or modify an existing one to enable SEV support. 
%Allocate sufficient resources (CPU cores, memory) to the VM.

\noindent\textbf{Results.} We consider 80\% of the entire dataset as training and validation data for the offline phase. The remaining 20\% are separated for testing in the online phase. A CNN model dedicated to the AMD SEV-based execution environment is trained with the training dataset. The training data contains $107 \times 60 = 6420$ measurements, where 60 measurements from each Docker image are randomly chosen. After getting a stable prediction accuracy on the validation data that contains 20 additional measurements from each Docker image, we fixed the pre-trained model to be used for the online phase. In the online phase, we tested the accuracy of our pre-trained model with $107 \times 20 = 2140$ measurements. The test accuracy for the AMD SEV execution environment is 79.1\%. Although we shifted our experiment from the simplest execution environment to the most secure one, we can observe that our attack still remains feasible.

% \subsection{Intel TDX}
% \paragraph{Experimental Setup}
% CPU: 2x Intel Xeon Platinum 8480CTDX, 2 sockets, 56 cores per socket, 2 threads per core
% TDX: TDX module version 1.0, build date: 2023-02-06, build num: 457
% Host kernel: 5.19.0 compiled to use TDX by default
% Host OS: Ubuntu 22.04.2 LTS
% Containerd, Kata, QEMU versions like with the SEV setup.
% VM: 1 vCPU
% Guest kernel: 5.19-TDX-v2.2-112-tdx
% \paragraph{Data Collection}
% \paragraph{Results}

%\subsection{Take-away message}

%\eduard{- Compare container technologies vs VM technologies}

%\eduard{- Compare container technologies without and with TEE}

%\eduard{- Compare VM technologies without and with TEE}
%\berk{less fingerprint from some Docker images}
\section{Evaluation}\label{sec:evaluation}
In \autoref{sec:case_study}, we showed that individual Docker containers create distinguishable CPU frequency fingerprints when run on top of different sandbox environments. This section performs additional analysis based on the previous results to make the threat model more practical. Specifically, we evaluate the portability of our attack (E1), the effect of sample size (E2), the simultaneous execution of multiple containers (E3), the relation between frequency activity and misprediction rates (E4), the detection of different image versions (E5), the effects of input arguments (E6) and multi-core execution (E7), noise analysis (E8) and effects of Docker pull on Firecracker (E9). %Afterward, we analyze the impact of our attack in a threat scenario where multiple containers are running concurrently. We also investigate the potential reason behind the high misprediction rate for some Docker images. In addition, we are interested in fingerprinting the MicroVM created by the firecracker VMM to know when we expect a user to pull or run any container inside the MicroVM. Pulling Docker images inside a MicroVM might cause unique fingerprints. Although a user might pull an image beforehand and run it later, it is still an important question to ask whether pulling an individual Docker image creates any specific changes in the CPU frequency.
%\eduard{(Eduard to check this again)}

\begin{comment}
\begin{table*}
\centering
\caption{Comparison of the fingerprinting results in different execution environments and microarchitectures. The highest accuracy for each environment is given with the bold font.}
\label{tab:compare_result}
  \begin{tabular}{cccccccc}
    \toprule
    \multirow{2}{*}{ \textbf {Execution Environment} }
        & \multirow{2}{*}{ \bfseries \# of containers }
        & \multicolumn{6}{c}{ \bfseries microarchitectures } \\ 
    \cmidrule{3-8} &     & Comet Lake           & Cascade              & Broadwell            & Skylake              & Coffee Lake          & EPYC                 \\
    \midrule
    Native         & 126 & {\bf 84.50\%}        & {83.03\%}            & {\color{red}73.37\%} & {\color{red}81.04\%} & {\color{red}74.16\%} & {\color{red}79.60\%} \\
    Firecracker    & 126 & -                    & {\bf 73.04\%}        & -                    & {72.01\%}            & -                    & -                    \\
    gVisor         & 126 & -                    & {71.20\%}            & {\bf 71.7\%}         & -                    & -                    & -                    \\
    Gramine        & 50  & -                    & -                    & -                    & -                    & {\bf 91.4\%}         & -                    \\
    AMD-SEV        & 107 & -                    & -                    & -                    & -                    & -                    & {\bf 79.8\%}         \\
    \bottomrule
  \end{tabular}
\end{table*}
\end{comment}

\begin{table*}
\centering
\caption{Comparison of the fingerprinting results in different execution environments and microarchitectures. The highest accuracy for each environment is given with the bold font.}
\label{tab:compare_result}
\begin{tabular}{|c|c|c|c|c|c|c|c|} 
\hline
\multirow{2}{*}{\begin{tabular}[c]{@{}c@{}}\textbf{Execution}\\\textbf{ Environment}\end{tabular}} & \multicolumn{1}{c}{\multirow{2}{*}{\begin{tabular}[c]{@{}c@{}}\textbf{\textbf{\# of}}\\\textbf{\textbf{containers}}\end{tabular}}} & \multicolumn{6}{|c|}{\textbf{Microarchitectures}}                     \\ 
\cline{3-8}
                                                                                                   & \multicolumn{1}{c|}{}                                                                                                               & Comet Lake & Cascade & Broadwell & Skylake & Coffee Lake & AMD EPYC  \\ 
\hline
Native                                                                                             & 126 & \textbf{84.5\%}     & 83.03\% &   73.37\%       &  81.04\%      &   74.16\%          &       79.60\%   \\ 
\hline
Firecracker                                                                                        & 126                         & -          & \textbf{73.04\%} & -         & 72.01\% & -           & -         \\ 
\hline
gVisor                                                                                             & 126                         & -          & 71.20\% & \textbf{71.7\%}    &       & -           &         \\ 
\hline
Gramine                                                                                            & 50                          & -          & -       & -         & -       & \textbf{91.4\%}      & -         \\ 
\hline
AMD-SEV                                                                                            & 107                         & -          & -       & -         & -       &      -       & \textbf{79.8\%}    \\ 
\hline
\multicolumn{8}{l}{\begin{tabular}[c]{@{}l@{}}
\\\end{tabular}}                                                                                              
\end{tabular}
\end{table*}

\smallskip
\noindent\textbf{\textit{E1) Compatibility with different microarchitectures and execution environments.}} 
The ideal environment for leveraging CPU frequency to identify different containers is the native Linux environment. Although containers are isolated through cgroups and namespaces, our side-channel attack is still effective in identifying running containers. From \autoref{tab:compare_result}, we can observe that the highest accuracy (84.5\%) is obtained from the native environment when the number of images is higher than 100. Proceeding toward more complex execution environments is expected to reduce the test accuracy due to additional noise. However, the increased level of security features and additional layers in different execution environments cannot mask the CPU frequency fingerprint of the containers. In \autoref{tab:compare_result}, the container profiling accuracy is enlisted for four additional secured execution environments. In all cases, the running containers can be detected with more than 70\% accuracy. It is to be noted that we included all the Docker containers from Docker Hub~\cite{DockerHubAPI} that work without any issue within the individual execution environment in our target devices. The rest of the Docker containers did not run due to issues specific to the execution environment and our target devices. 

Previously, the efficacy of our attack was tested on a specific microarchitecture. In this section, we evaluate the performance of this attack in different microarchitectures to prove the attack's portability. It is to be noted that the portability indicates the reproducibility of the attack in a different microarchitecture rather than the transferability of the pre-trained model from one microarchitecture to another. %In this context, each previous execution environment is created in different microarchitectures, and the same experiment is carried out to analyze the performance. 
As shown in Table~\ref{tab:compare_result}, the fingerprinting attack is compatible even with different microarchitectures for a specific execution environment. The detection accuracy remains close even though a different microarchitecture is used with the same execution environment. For every execution environment, we consider testing our attack in two different microarchitectures except for Gramine and AMD-SEV, as our other available servers do not support trusted execution environments, like SGX and SEV.
%Continued..

\begin{comment}
\begin{table}[ht!]
\begin{tabular}{l|ccll}
\hline
\multicolumn{1}{c|}{\multirow{2}{*}{\textbf{\begin{tabular}[c]{@{}c@{}}Execution\\ Environment\end{tabular}}}} & \multicolumn{4}{c}{\textbf{microarchitectures}}                                                                                                          \\ \cline{2-5} 
\multicolumn{1}{c|}{}                                                                                          & \multicolumn{1}{c|}{Comet Lake} & \multicolumn{1}{l|}{Cascade} & \multicolumn{1}{l|}{Broadwell} & Skylake             \\ \hline
Native                                                                                                & \multicolumn{1}{c|}{84.5\%}              & \multicolumn{1}{c|}{83.03\%}          & \multicolumn{1}{l|}{}                   &                              \\ \hline
Firecracker                                                                                            & \multicolumn{1}{l|}{}                    & \multicolumn{1}{c|}{73.04\%}          & \multicolumn{1}{l|}{}                   & \multicolumn{1}{c}{72.01\%} \\ \hline
gVisor                                                                                                 & \multicolumn{1}{l|}{}                    & \multicolumn{1}{c|}{71.2\%}           & \multicolumn{1}{c|}{77.8\%}             &                              \\ \hline
\end{tabular}
\end{table}
\end{comment}

\begin{figure}[t!]
    \centering
    \includegraphics[width=\linewidth]{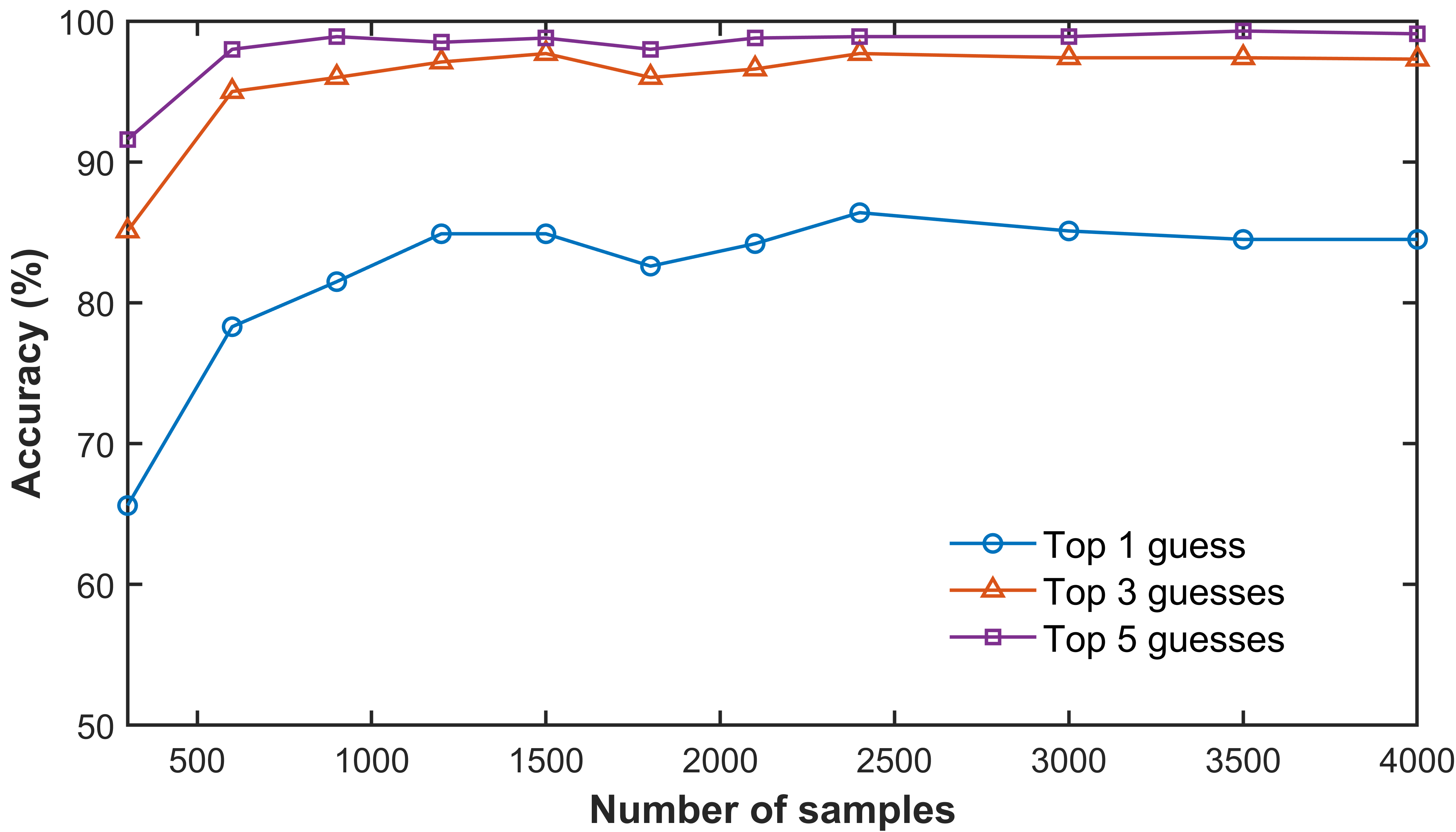}
    \caption{Accuracy of container fingerprinting in the native environment with different samples. The accuracy in terms of the top 3 and top 5 guesses of the prediction model is also considered to make a comparison. %It is to be noted that the top 1 guess refers to the original prediction of the ML model with the highest confidence rate.
    }
    \label{fig:native_accuracy}
    \vspace{-5mm}
\end{figure}

\smallskip
%\newpage
\noindent\textbf{\textit{E2) Impact of varying sampling size in the accuracy of the ML model.}}
%We are interested in evaluating the result from the perspective of the sample size of the fingerprints. 
Although we initiated the experiment with 4000 samples per measurement, we also analyzed the impact of different sample sizes with a fixed resolution to select the least amount of samples that are adequate to perform the fingerprinting with satisfactory accuracy. Reducing the sample size will speed up the overall data collection and attack execution time, making the attack more efficient and stealthy. We perform this experiment on the Native Linux environment. In \autoref{fig:native_accuracy}, we observe that the accuracy does not improve further after 2500 samples, resulting in a 1.7 times faster attack. The test accuracy of the pre-trained model with 2500 samples is 86.4\%, which is even slightly better than 4000 samples (84.5\%). One of the reasons for such a scenario is that we observe more activity during the starting phase of the Docker runtime, and it facilitates the ML model to extract more useful features from the initial samples. The top guesses of the prediction model are also analyzed to explain the deviation of the misprediction.% We determined the top 3 guesses and top 5 guesses of the pre-trained model for our test data. 
The accuracy of the pre-trained model with top 3 and top 5 guesses are 97.3\% and 99.1\%, respectively, with 4000 samples. A similar trend can be seen in \autoref{fig:native_accuracy} with the reduced number of samples. The accuracy with the top 3 and top 5 guesses remains stable after 2500 samples. 

\smallskip

\noindent\textbf{\textit{E3) Effects of running multiple containers simultaneously.}}
In cloud environments, multiple tenants share the same physical server simultaneously, where each tenant's workload may be run on one specific core. In a scenario where multiple tenants are running Docker images on their assigned cores at the same time, the efficacy of our attack from the adversary's perspective is evaluated in this section. We intend to measure the effects of running multiple containers simultaneously in separate cores, where each container is pinned to a specific core. Since the \texttt{cpufreq} subsystem reports the dynamic frequency of each core separately, an adversary can track multiple users or containers concurrently, which makes our attack stronger than other side-channel attacks in which only one victim can be monitored~\cite{gulmezoglu2017perfweb,aldaya2019port}. However, separate containers running in multiple cores might introduce some noise, creating fluctuation in core frequency and degrading the performance of the container detection model. We evaluate the performance degradation of the detection model in case an adversary monitors multiple physical cores concurrently, as well as the impact of running separate containers on two sibling threads.

\begin{figure}[t!]
    \centering
    \includegraphics[width=\linewidth]{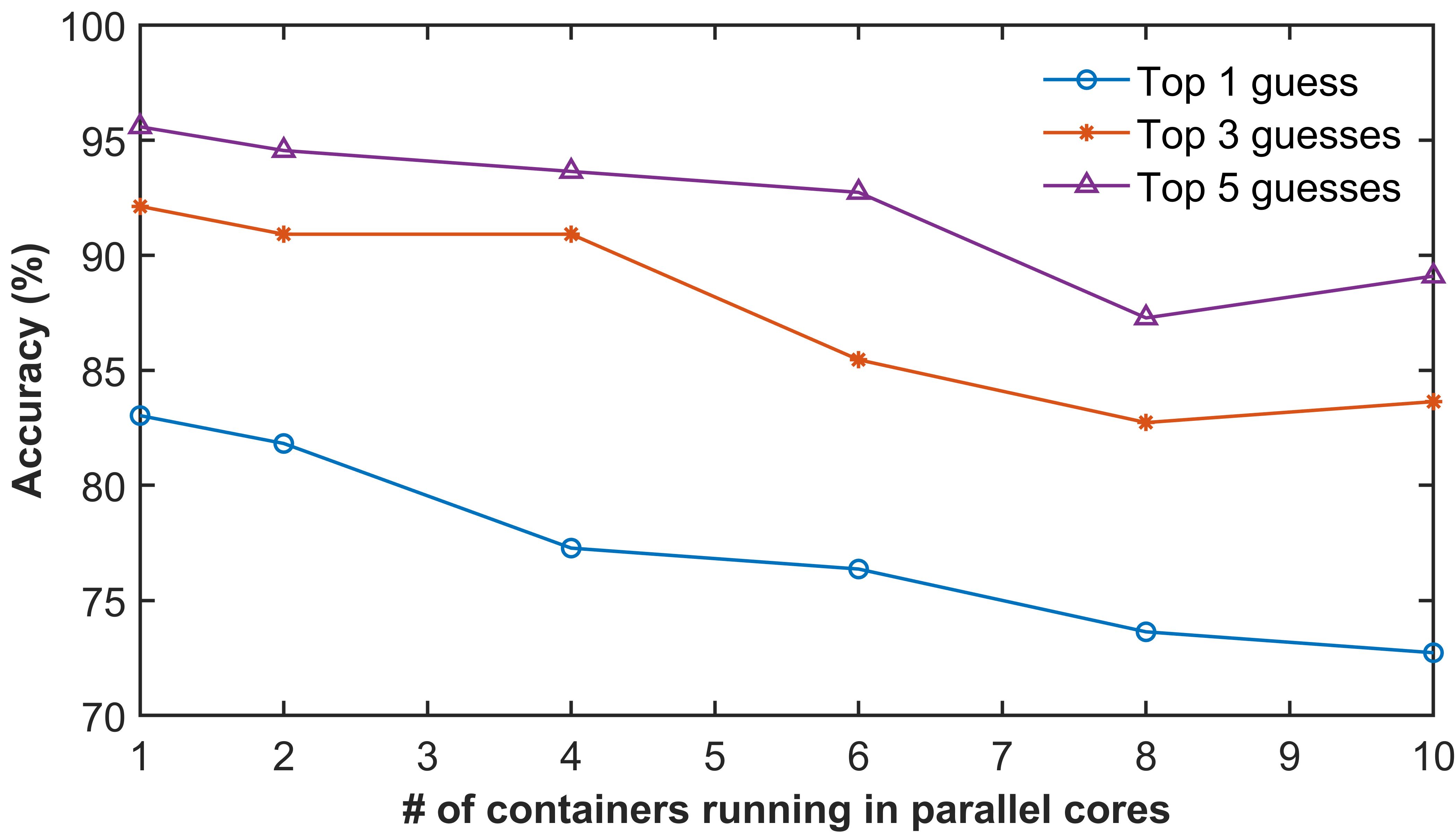}
    \caption{The effects of running containers simultaneously in parallel cores. Although the containers are running in separate cores, the concurrent execution introduces some noise, which affects the accuracy of the pre-trained model.}
    \label{fig:native_accuracy_parallel_run}
    \vspace{-5mm}
\end{figure}

This experiment is performed on the Intel Cascade microarchitecture with the native environment. In order to analyze the impact of noise, we randomly choose a set of containers and run them in two separate cores in parallel. By monitoring the frequency values from these two cores, we collected 110 measurements and fed them to our pre-trained model, which was previously trained in a noise-free environment in the Intel Cascade microarchitecture. The accuracy based on our newly collected test data is 81.8\% in contrast to the accuracy of 83.03\% obtained with noise-free test data. When we increase the number of containers to four, the accuracy drops to 77.3\%. We repeated the experiment by running 6, 8, and 10 containers in parallel to realize how far it affects the accuracy. In Figure~\ref{fig:native_accuracy_parallel_run}, we demonstrate the accuracy drop with the increasing number of containers in separate cores. By running ten containers concurrently in ten different cores, we can observe an accuracy drop of 12.4\% from a single container scenario. On average, every container introduces $\approx 1.5\%$ accuracy drop due to the added noise caused by concurrent execution. 
The top 3 and top 5 guesses predicted by the pre-trained model are also reported in Figure~\ref{fig:native_accuracy_parallel_run}. The accuracy of the pre-trained model based on the top 5 guesses is 95.6\% (noise-free), which drops to 89.1\% with ten simultaneous executions of containers. The margin of accuracy drop in this case is 6.8\%, compared with the accuracy drop of 9.2\% for the top 3 guesses and 12.4\% for the top 1 guess. It is to be noted that we can observe a rise in the accuracy with ten concurrent runs of containers for the top 3 and top 5 guesses. Since the containers running in parallel are picked randomly, there is a small fluctuation in detection accuracy while changing the number of cores. 

Another aspect of this experiment involves running two containers in the sibling threads of the same core. Interestingly, running them in the sibling threads reduces the accuracy to 42.7\%. This concludes that although it is still practical to fingerprint containers running concurrently on different cores, it is less effective if they run on the sibling threads of the same core. The reason is that two sibling cores affect each other's frequency values as they share many microarchitecture components, corrupting the fingerprint substantially. The impact of running containers on the same thread of a core is further analyzed in Appendix~\ref{appendix:1}.

\begin{figure}[t!]
    \centering
    \includegraphics[width=\linewidth]{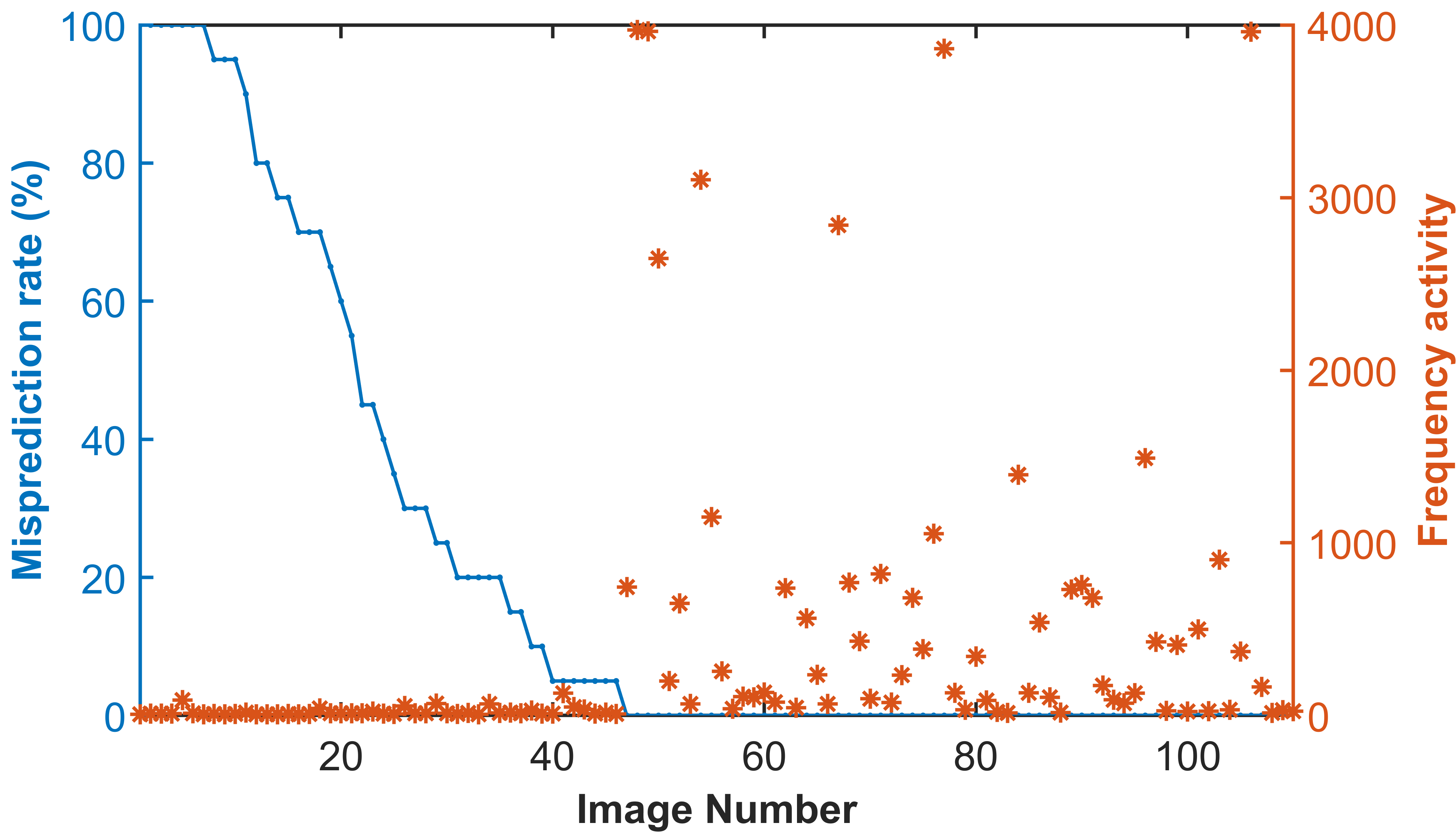}
    \caption{Demonstrating less frequency activity as the cause behind high misprediction rates of some Docker images. %The left y-axis represents the misprediction rate/image in descending order, while the right y-axis shows the corresponding frequency activity. 
    The left portion of the figure corresponds to the images that are comparatively harder to distinguish than the images in the right portion that show high-frequency activity.}
    \label{fig:activity_vs_misprediction}
    %\vspace{-5mm}
\end{figure}

\smallskip
\noindent\textbf{\textit{E4) Frequency activity vs misprediction rate.}}
For different execution environments, we achieve an accuracy of over 70\% in general. We put an effort into understanding the cause behind the mispredictions. In Figure~\ref{fig:activity_vs_misprediction}, the blue dots refer to the misprediction rate per image, sorted in descending order. Out of 110 Docker images, we find a 100\% misprediction rate for 8 images and a 0\% misprediction rate for 62 images. This indicates that some Docker images are more challenging to distinguish than others, affecting the overall accuracy of our prediction model. %After tracing back the frequency signatures of all of these images, we observed an interesting insight. 

We consider that the number of high-frequency points beyond a predefined threshold is considered as the frequency activity demonstrated by the runtime of a specific Docker image inside a container. According to our observation, Docker images that demonstrate less frequency activity have a higher misprediction rate compared to the others. Figure~\ref{fig:activity_vs_misprediction} illustrates this observation more comprehensively. The right y-axis of the figure attributes the frequency activity of individual images. On the left side of the figure, we can observe comparatively high misprediction rates with less frequency activity. However, on the right side, the frequency activity/image is comparatively higher, and the prediction model can distinguish the images with higher accuracy, thus causing zero misprediction rates. It is to be noted that this experiment is performed on the Cascade microarchitecture with the native environment. The minimum CPU frequency of this microarchitecture is 1 GHz; therefore, we set our predefined threshold to be 1.2 GHZ (adding a small factor with the minimum frequency to incorporate the effect of small noise) for quantifying the frequency activity for each runtime of the Docker images.

\begin{figure}[t!]
    \centering
    \includegraphics[width=\linewidth]{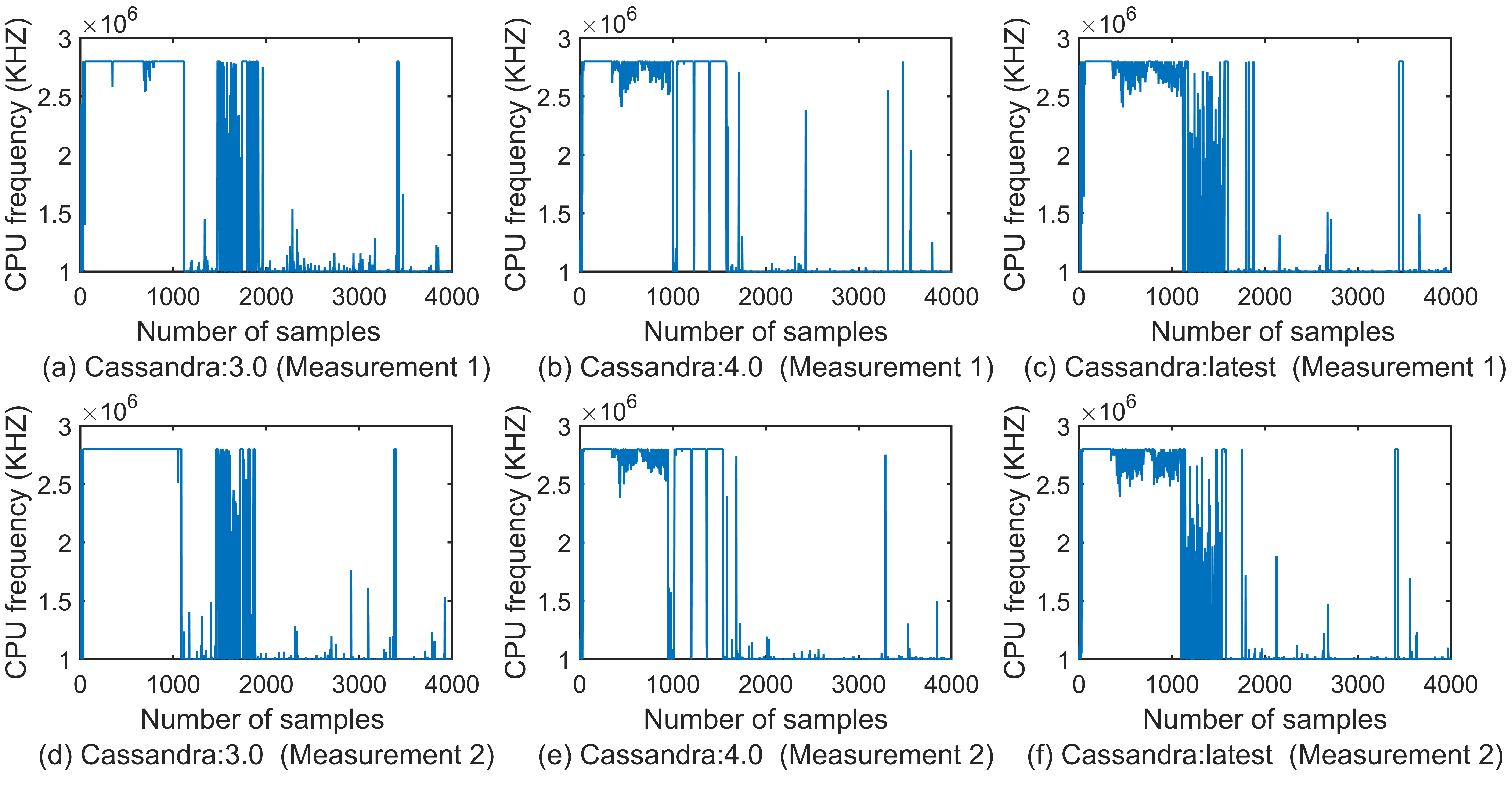}
    \caption{The CPU frequency signatures of three different versions of Cassandra image. Different versions of the same image exhibit unique variations in the CPU frequency signature that make them distinguishable by the DL model. Two measurements for each version are presented to show the consistency of the pattern.}
    \label{fig:version_image}
    \vspace{-5mm}
\end{figure}

\smallskip
\noindent\textbf{\textit{E5) Feasibility of the attack with different versions of the docker images.}}
The previous experiment considers the latest versions of the docker images. However, the victim can run different versions of an image, leading to variations in the image layers, such as dependencies. We carry out an experiment to analyze the CPU frequency signature for different versions of the docker images. For this analysis, we randomly selected 25 images from our image list and chose five different versions for each image. Our initial observations imply that different image versions have unique signatures, as shown in Figure~\ref{fig:version_image}. Although there are common patterns among different versions of the same image, a well-trained ML/DL model can distinguish versions with high accuracy. To prove our hypothesis, we collected 50 measurements (CPU frequency signatures) for all five versions of the 25 images. By considering each of these versions as separate classes, we trained a CNN model with 60\% of the training data and 20\% validation data. The rest 20\% data are used for testing the pre-trained model in the online phase. The acquired test accuracy for the pre-trained model is 81.02\%, demonstrating that different versions of Docker images produce adequate variability in signatures to fingerprint them. 

\begin{figure}[t!]
    \centering
    \includegraphics[width=\linewidth]{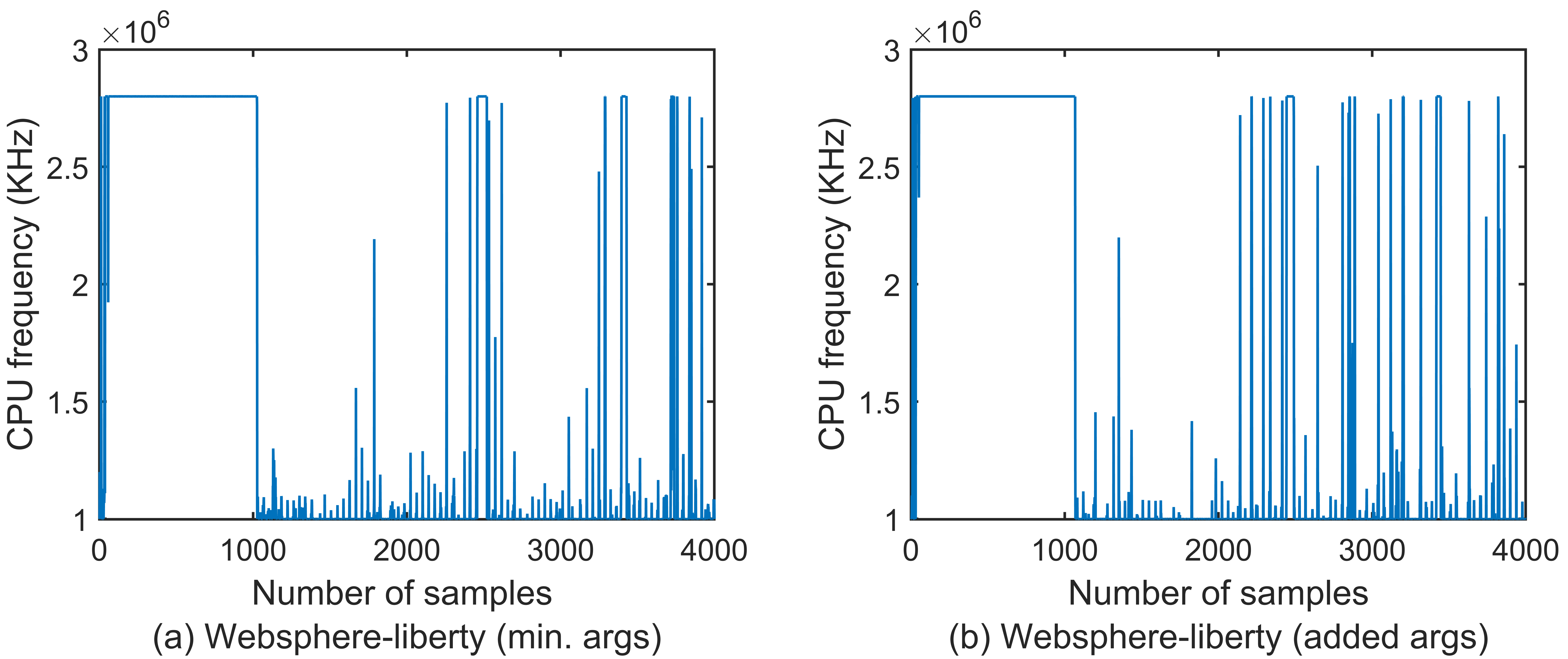}
    \caption{The CPU frequency signatures of \texttt{websphere-liberty} run with two sets of arguments, where (a) corresponds to the set with minimal arguments and (b) refers to the set with additional arguments. We observe negligible changes in the fingerprints.}
    \label{fig:diff_args}
    \vspace{-5mm}
\end{figure}

\smallskip
\noindent\textbf{\textit{E6) Effects of input argument variations in docker images.}}
For a more realistic end-to-end attack scenario, we considered more flexibility from the victims' side. The victim can run different docker images with variations in the input arguments. Our attack considers these aspects and has been tailored to accommodate these variations. %As explained in Section~\ref{sec:evaluation}, different versions of a docker image create unique CPU frequency signatures. Hence, the adversary can train a model with the CPU frequency signatures of different versions of the docker images in the offline phase to predict the image name with its version. 
The proposed attack is tested with the variation of input arguments. We randomly selected 20 images and collected CPU frequency data while running them with two different sets of arguments (Appendix~\ref{appendix:8}). One of the sets includes minimal arguments to run the image, and the other set contains more arguments specific to the image. When we considered more arguments to run the images, we noticed no significant variation in the signature. In Figure~\ref{fig:diff_args}, the signatures of the same image run with two sets of arguments are compared. We observed minor changes in the later part of the signature, but these did not affect the classification accuracy. The initial portion of the signature remains almost similar, which makes the image distinguishable even with additional inputs.

\smallskip
\noindent\textbf{\textit{E7) Effects of assigning containers to multiple cores.}}
We analyzed the feasibility of our attack when a docker image is assigned to multiple cores. If we restrict the docker image to run in two separate cores, the workload gets distributed between the two cores. Consequently, the active frequency signature of the docker image gets divided into two parts, which shrinks the active area of the fingerprint while having a similar pattern in both cores. We also observed that fingerprints are more noisy due to frequent jumps between cores, and we used a Gaussian filter with a window size of 10 samples to eliminate the noise. Afterwards, we performed the \textit{movmax} operation with the same window size to smooth out the signal. We randomly chose 45 images and initially restricted them to utilize only two cores during their runtime. We collected 100 measurements of CPU frequency signatures from both of these active cores for each image and filtered out the noise. As we observed similar patterns in the signatures from both cores for a single image, instead of combining them, we considered each signature as a separate measurement for that image. After training a model with the newly collected fingerprints, we achieved an accuracy of 73.1\% from 45 images. 

In the second stage, we restricted the images to four cores. After filtering out the noise and smoothing the signal in the same manner as before, we observed a similar pattern in all four cores for the same image, while the active portion of the signature decreases almost by half in each core compared to the two-core scenario. Considering the signatures from each core as separate measurements, we trained a new CNN model only for the 4 core scenario. The accuracy decreased to 68.9\% for the same 45 images. If we increase the number of shared cores, each time, the active portion of the signatures gets reduced in every active core as the workload is distributed among them. Considering the top 3 and top 5 guesses of the pre-trained model in the two cores scenario, we achieve an accuracy of 86.44\% and 90.22\%, respectively. If the docker images share 4 cores, we achieve 78.33\% and 84.81\% accuracy considering the top 3 and top 5 guesses, respectively. With an enhanced number of shared cores, relying on the top 3 or top 5 predictions from our pre-trained model is essential for efficiently narrowing down our search for fingerprinting the correct docker image.

\begin{figure}[t!]
    \centering
    \includegraphics[width=\linewidth]{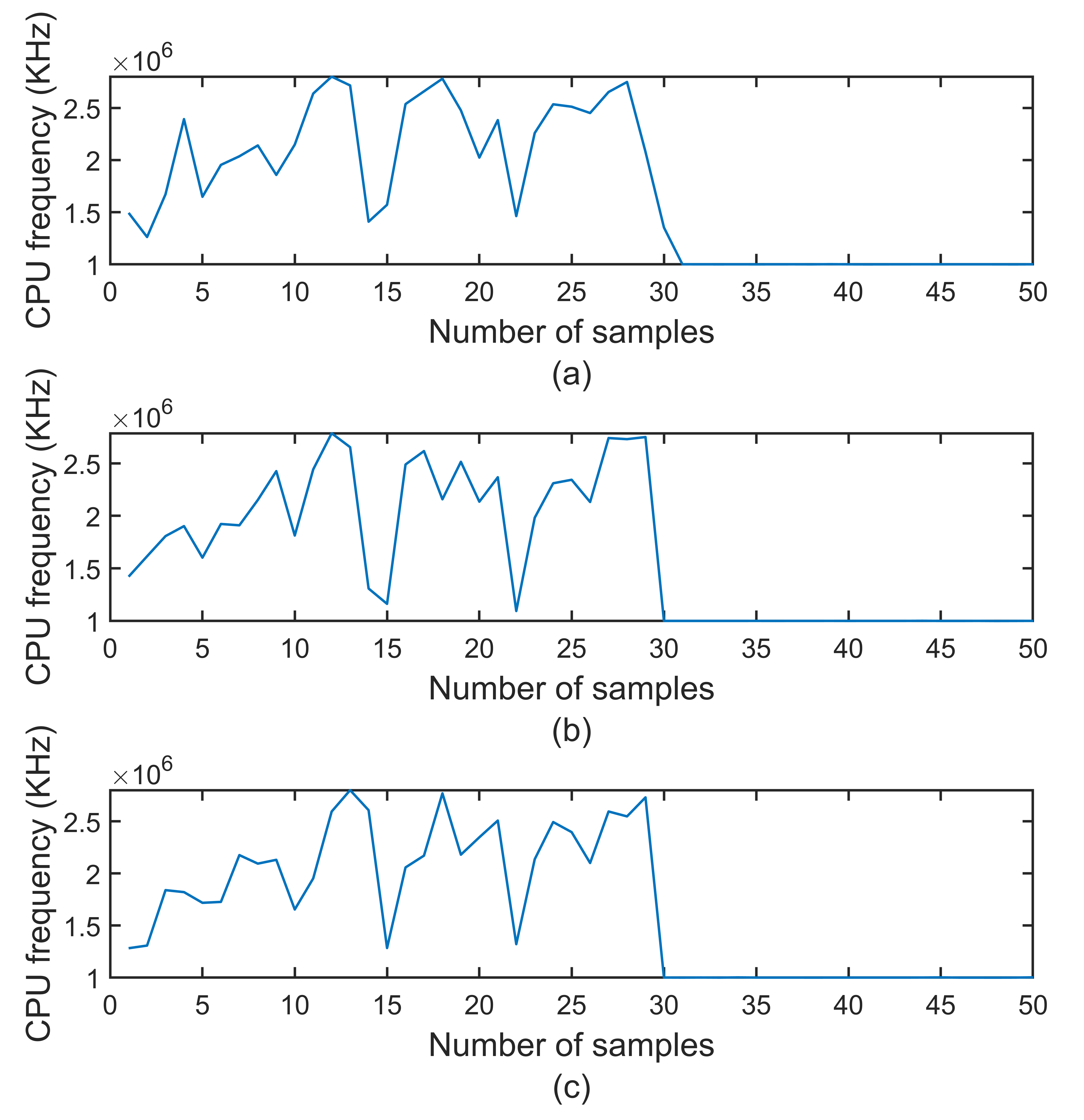}
    \caption{CPU frequency fingerprints for the bootstrap phase of firecracker-containerd (for Firecracker VMM).}
    \label{fig:boot}
    \vspace{-5mm}
\end{figure}

\smallskip
\noindent\textbf{\textit{E8) Noise analysis on Firecracker VMM.}}
The initiation of the Firecracker execution environment starts with the bootstrap phase of the firecracker-containerd. The \texttt{firecracker-containerd} is a customized containerd binary that runs in the background to initialize the containerd runtime as explained in the previous section. This bootstrap phase can be considered as the first form of noise directly associated (tied) with the Firecracker execution environment.  
The frequency signature of the bootstrap phase is shown in \autoref{fig:boot}. We collect the frequency fingerprints while starting the bootstrap phase in parallel. In the figure, we present three measurements corresponding to the same noise. As the bootstrap phase occurs fast, a rise in the CPU frequency can be observed by analyzing the first few samples. Once the bootstrap is finished, the frequency returns to the base frequency. It is good to see that the frequency signatures remain similar for each individual measurement. This provides an opportunity for the adversary to figure out the moment when the victim starts using the Firecracker environment.

%The starting of MicroVM creates a specific signature which We have found to be repeated in every measurement of the data that we collected during the runtime of the containers. \autoref{fig:start_MicroVM} demonstrates the fingerprints corresponding to multiple instances of starting of the MicroVM. The figure shows that the fingerprints remain similar during every startup of the MicroVM. \berk{Interpret Figure 4 and 5 in text. This is poor writing.}

\begin{figure}[ht!]
    \centering
    \includegraphics[width=\linewidth]{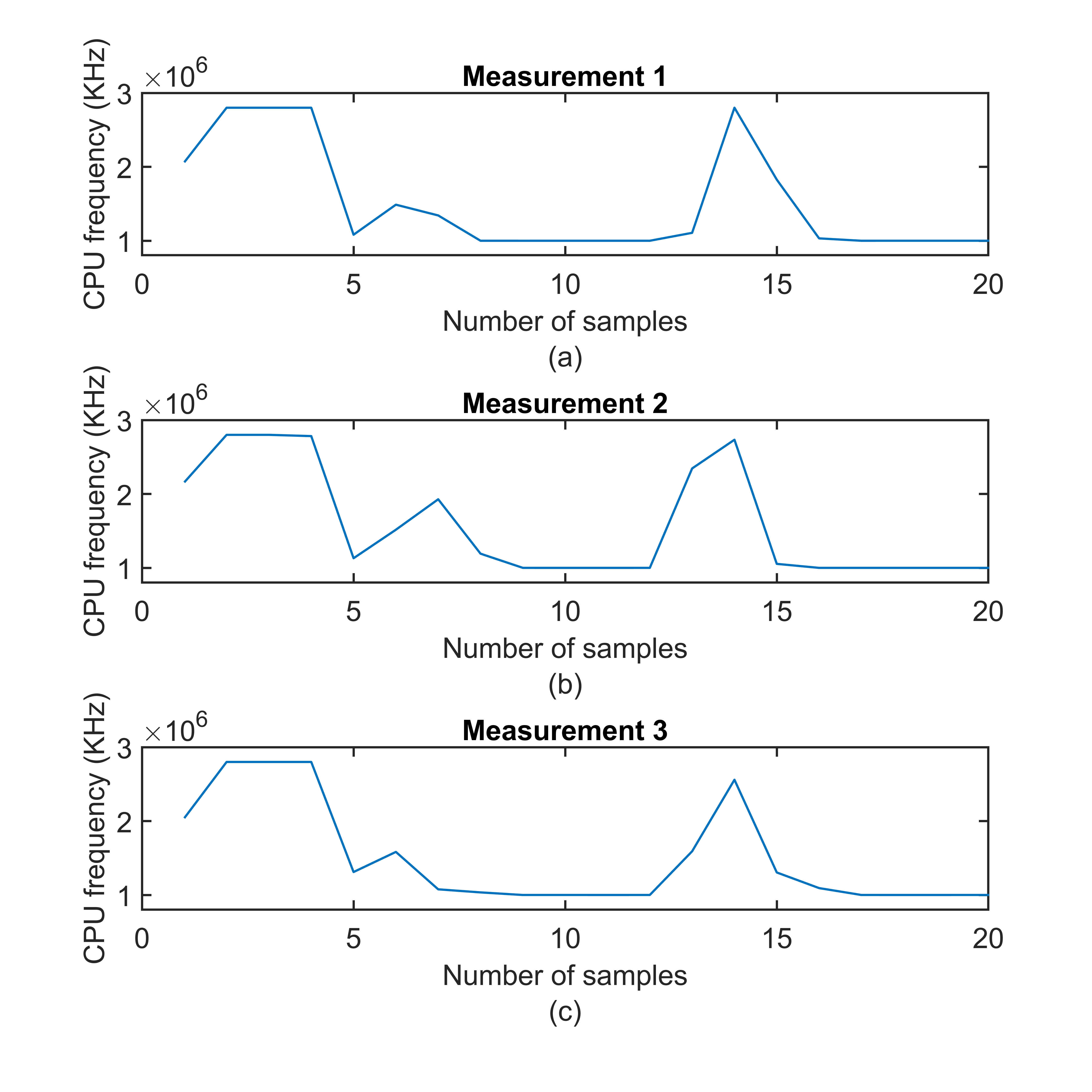}
    \caption{Induced noise due to the starting of a MicroVM (for Firecracker VMM). The frequency signatures show similar variation in all three measurements.}
    \label{fig:start_MicroVM}
\end{figure}

Once the bootstrap phase is complete, the \texttt{firecracker-ctr} tool pulls all the necessary images. This command line tool also associates the \texttt{aws.firecracker} runtime to set forth a request to start a MicroVM and run any pulled image inside that MicroVM. The starting phase of the MicroVM introduces an additional noise in our frequency signature data. In the threat model (\autoref{sec:threat_model}), we mentioned that each MicroVM only runs one container inside. Therefore, the frequency signature of every Docker container will contain some initial noise created from the starting MicroVM part. Appendix~\ref{appendix:2}:\autoref{fig:start_MicroVM} represents the signature of starting the MicroVM for three different instances. We expect that the start-up process of the MicroVM should involve similar activity and affect the CPU frequency in the same manner. The CPU frequency signatures in Appendix~\ref{appendix:2}:\autoref{fig:start_MicroVM} demonstrate a similar pattern as we expected.

\begin{comment}
\begin{figure}[t!]
    \centering
    \includegraphics[width=\linewidth]{Figures/start_MicroVM.png}
    \caption{Induced noise due to the starting of a MicroVM (for Firecracker VMM). The frequency signatures show similar variation in all three measurements.}
    \label{fig:start_MicroVM}
\end{figure}
\end{comment}

\smallskip
\noindent\textbf{\textit{E9) Effects of Docker pull on Firecracker VMM.}}
%We examine the bootstrap phase of the \texttt{firecracker-containerd} process to determine whether an adversary can detect the starting time of the container run.% as it provides sufficient information to the adversary to know whenever the victim is going to utilize the Firecracker tool.
%For this purpose, frequency values are monitored for the bootstrap phase as given in \autoref{fig:pull_vs_run}.
Our ML model trained with the fingerprints collected from the Firecracker environment categorizes the Docker images based on their runtime signatures (i.\,e., after running the \texttt{docker run} command). However, the fingerprints do not include the pulling phase of the Docker images since the pulling phase is generally performed once before running the container. On the other hand, we consider a more interesting scenario in which the victim might pull the image instantly before running the container and might remove the image after the utilization. In this scenario, the attacker collects a fingerprint that captures frequency values from both the pulling phase and execution portion of the Docker image. In order to determine whether it is possible to separate these two phases, we first collect fingerprints for the pulling phase of different Docker images inside the MicroVM. When we analyze these fingerprints as visualized in Appendix~\ref{appendix:3}: \autoref{fig:pull}, we observe that the pulling phase of each layer in Docker images creates a distinguishable frequency signature because the number of layers and their contents in each image are distinct. For example, we observe high-frequency patterns up to 2200 samples for \texttt{Erlang}. However, for \texttt{Percona}, this pattern exists for 1000 samples. In the same figure, we also present two different measurements of the pulling phase of the same image. It is to be noted that we removed the previously pulled image before the collection of the new measurement to determine whether the signature remains consistent over multiple measurements. 
\begin{comment}
\begin{figure}[t!]
    \centering
    \includegraphics[width=\linewidth]{Figures/pull_comparison.png}
    \caption{Comparison of pulling different images inside a MicroVM of Firecracker. (a, b) represent two different measurements of pulling the \texttt{Erlang} image, and (c, d) refers to two separate measurements for pulling the \texttt{Percona} image. As can be seen, it is possible to recognize the container image being pulled in the Firecracker execution environment based on the CPU frequency signatures.}
    \label{fig:pull}
\end{figure}
\end{comment}

\begin{figure}[t!]
    \centering
    \includegraphics[width=\linewidth]{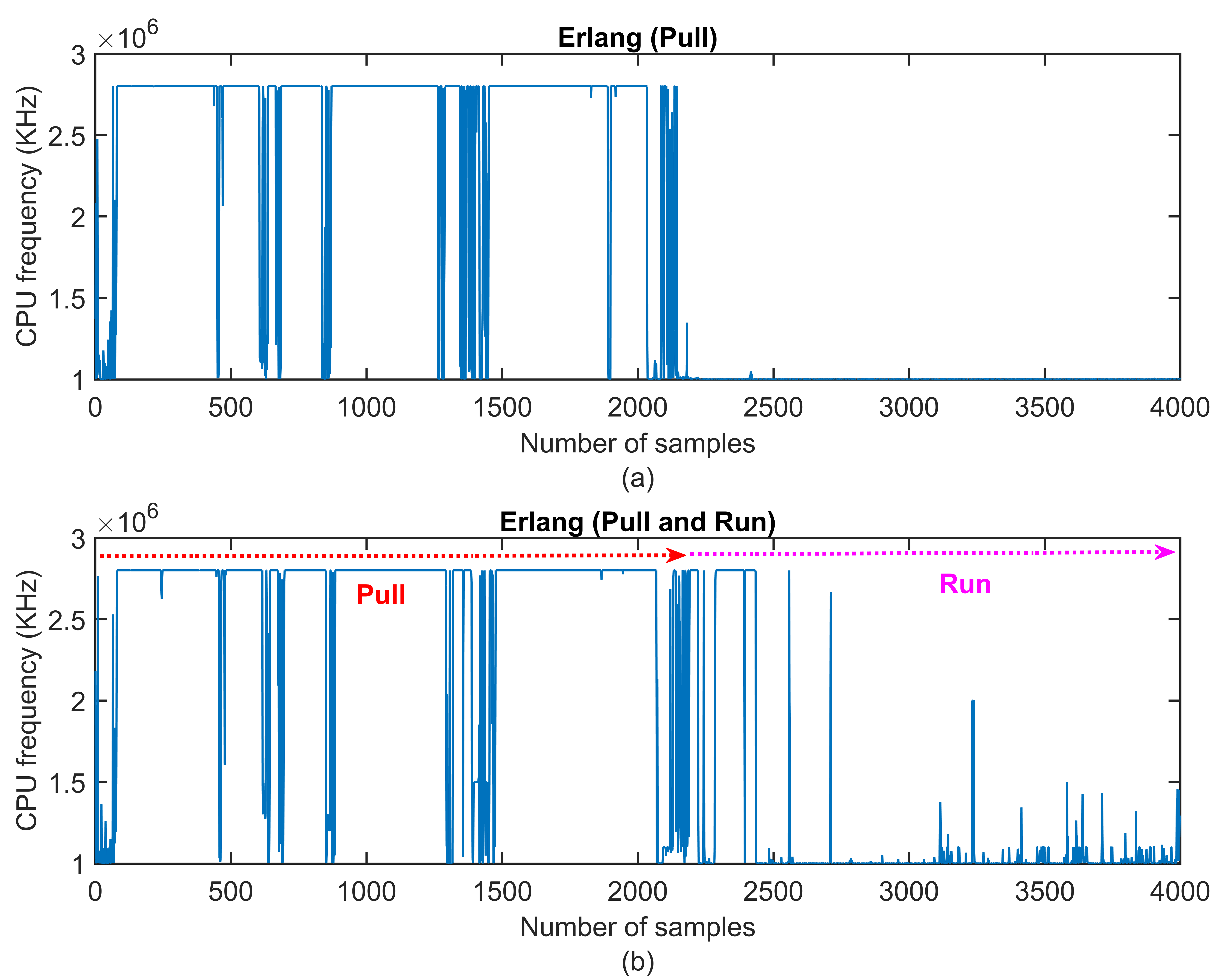}
    \caption{Fingerprints corresponding to the start of a MicroVM (for Firecracker VMM).}
    \label{fig:pull_vs_run}
    \vspace{-5mm}
\end{figure}

Afterward, we collect CPU frequency while pulling and running the image together. 
In \autoref{fig:pull_vs_run}, the CPU frequency fingerprint of pulling vs pulling and running the \texttt{Erlang} container is shown. \autoref{fig:pull_vs_run}(a) shows the measurement when we only pulled the \texttt{Erlang} image inside the MicroVM. Conversely, \autoref{fig:pull_vs_run}(b) presents the CPU frequency signature of a new measurement where we pull and run the \texttt{Erlang} image together. By comparing \autoref{fig:pull_vs_run} (a) and (b), we can observe that it is possible to recognize the pulling part of the Docker image as it remains consistent. There can be an issue regarding the alignment, which can be solved based on the signature of the MicroVM start operation, shown in Appendix~\ref{appendix:2}:\autoref{fig:start_MicroVM}. Thus, an adversary can understand different operations based on the CPU frequency data by conducting a more detailed analysis of the data.

\section{End-to-end Attack}

Adversaries are always driven to execute attacks that inflict significant damage while minimizing costs and the likelihood of detection as much as possible. Despite the prevalence of Docker images containing outdated packages with high-severity vulnerabilities, launching an attack “blindly“ reduces the chances of success and increases the risk of detection. Our attack can be viewed as a means of acquiring information that enables the execution of more effective, efficient, and stealthy attacks at a late stage.

The end-to-end attack presents various scenarios for consideration from the adversary’s perspective. Adversaries may choose to execute a targeted fingerprinting attack against a specific docker image or instead target any docker image running on the server where they are located. Irrespective of their objective, our attack always comprises an offline phase followed by an online phase. In its simplest form, during the offline phase, the adversary first compiles a dataset by collecting frequency measurements from accessible docker images and then utilizes this dataset to train a machine learning model, with each image representing a distinct class within the model. Note that the training and storage of the machine learning model can occur on a separate device accessible only to the adversary. Importantly, adversaries can accurately replicate the environments where cloud containers operate. This is because all information about sandboxing mechanisms and microarchitectures employed by various cloud providers is publicly available online~\cite{aws_instance, aws_amd, alibaba_ecs, amd_google_cloud}. Once the adversary executes the attack and identifies the Docker image in use, they can leverage information from security databases to identify vulnerabilities affecting said image and conduct a more targeted attack. In Section 9, the feasibility of fingerprinting the docker images with different versions, input arguments, and multiple cores/images is demonstrated.

In certain scenarios, additional details must be considered. For instance, when the adversary intends to execute a targeted fingerprinting attack from a malicious container (or similar sandbox), they may employ various strategies to increase the likelihood of achieving co-location with the victim docker image~\cite{DBLP:conf/ndss/FangWNOSKRH22, inci2016co}. Another aspect to consider is the effort the adversary must put into the offline phase. One possibility for the adversary is to train a different ML model for each sandbox – microarchitecture pair in the cloud provider where the victim runs. Given that the number of microarchitectures and sandboxes utilized by cloud providers is relatively small and that the offline phase happens very occasionally, we believe that this approach is reasonable. However, if the adversary possesses information regarding the victim image and its requirements, they can narrow down their options and develop a smaller set of ML models. Finally, we expect adversaries can employ the state-of-the-art available techniques in ML to build models that can detect docker images that were not included in the training dataset or to update the generated ML models without needing to start the training from scratch each time.

\section{Related Work}\label{sec:related_work}
In this section, we provide a brief overview of related work in side-channel attacks in cloud computing and OS-based side-channel attacks.

\subsection{Side-channel Attacks in Cloud Computing}

Cloud computing has been a target for side-channel attacks since it was shown that co-location between users can be detected with a high accuracy~\cite{ristenpart2009hey}. After this study, side-channel attacks on cloud computing focused on shared resources between tenants. Zhang et al.~\cite{zhang2012cross} showed that RSA decryption keys can be leaked through cache attacks. Next, Inci et al.~\cite{inci2016cache} showed that co-location detection is still possible on Amazon EC2 clouds, and last-level cache usage can be monitored to reveal 2048-bit RSA keys belonging to users. The deduplication feature utilized in cloud computing was also leveraged to leak AES keys from co-located users~\cite{irazoqui2014wait}. The memory bus is also targeted to create extensive contention on the victim's application, resulting in slowing down the system~\cite{wu2014whispers}. Delimitrou et al.~\cite{delimitrou2017bolt} showed that resource pressure of the co-located VMs reveals the victim application with high accuracy. Similarly, Gulmezoglu et al.~\cite{gulmezoglu2017cache} demonstrated that last-level cache usage can be used to detect running applications in co-located VMs on the EC2 cloud. Potential convert and side-channels are also examined in cloud computing in which multiple containers are run at the same time~\cite{gao2017containerleaks}. This study shows that power outage attacks can be performed by co-resident containers in cloud servers.

%\subsection{Other Attacks in Cloud Computing}
%Even though microservices and serverless computing platforms are relatively newer and still in the progress of development, several attacks have been presented on these platforms. Open Web Application Security Project (OWASP) published a report on the security risks of serverless computing platforms~\cite{owasp}, stating that the design of serverless platforms enables different types of attacks such as event injection~\cite{jeremydaly,o2020serverless,krug_blackhat}, misconfigured authentication mechanisms and broken access control~\cite{krug_blackhat,segal_iam,rich}, Denial of Service (DoS) and Denial of Wallet (DoW) attacks~\cite{kelly2021denial,wang2018peeking}, cross-site scripting (XSS)~\cite{marin2022serverless}, and insufficient logging and monitoring.

\subsection{OS-based Side-channel Attacks}

Several built-in sensors can be accessed through the OS features or interface, which were used to perform side-channel attacks, leaking private information from users. Zhang et al.~\cite{zhang2009peeping} leverages the \textit{procfs} system calls for keystroke detection. This attack shows that inter-keystroke timings can be recovered with high accuracy. Gulmezoglu et al.~\cite{gulmezoglu2017perfweb} used performance counters through the \textit{perf} interface available in Linux OS to track the website activity of a victim. Similarly, memory consumption statistics in resource tracking APIs are leveraged to track the usage of GPU cards to detect keystrokes and website activity~\cite{naghibijouybari2018rendered}. The power consumption measurable via \textit{MSRs} is also used to leak secret keys~\cite{lipp2021platypus} and website activity~\cite{zhang2021red}, showing the efficiency of power consumption sensors. Finally, dynamic frequency readings through the \textit{cpufreq} interface are used to leak AES-NI keys~\cite{wang2022hertzbleed} and website activity~\cite{dipta2022df}. ThermalBleed~\cite{kim2022thermalbleed} also exploits the thermal sensors to create covert channels and bypass kernel address space layout randomization (KASLR). Finally, Hot Pixels~\cite{taneja2023hot} shows that GPU and ARM SoCs are subject to thermal, power, and frequency attacks.

\section{Countermeasures}\label{sec:countermeasures}

%To defend against the fingerprinting attacks we identified, 

In this section, we propose an artificial noise injection mechanism and describe additional potential countermeasures to mitigate our attack. %To introduce them, we go from the simplest (and potentially less effective) to the more complex ones that offer stronger protection.

\smallskip
\noindent\textbf{Artificial Noise Injection.}  
We implemented a defense technique that injects random noise into the attacker's CPU frequency readings by increasing the workload in a controlled manner during the runtime of a docker image. We first assume a scenario where the attacker does not have knowledge regarding the noise injection tool. In the second scenario, we consider that the attacker has access to the noise injection tool and can re-train the ML model with noisy fingerprints.

To change the workload, we run a program that executes a pre-defined set of instructions iteratively in the sibling core as given in Algorithm~\ref{alg:noise}. In the noise injection code snippet, we utilize floating-point instructions as they involve complex mathematical operations on decimal numbers that require comparatively more computational resources than integer or logical operation instructions. Therefore, executing floating-point instructions increases the CPU frequency significantly compared to other instructions. Moreover, we observe that the core frequency increases further when different types of floating-point instructions are included in the noise injection code snippet. For example, we consider \texttt{fld} (load), \texttt{fstp} (store), \texttt{faddp} (addition), and \texttt{fmul} (multiplication) instructions and observe the frequency signature for individual instructions. It is to be noted that each instruction is executed 20 million times to increase the CPU frequency close to the maximum frequency level (approximately 2.8 GHz). To introduce more randomness into the noise, we use all these four instructions together to create a noise injection program as shown in Appendix~\ref{appendix:7} Algorithm~\ref{alg:noise}, which increases frequency based on the number of iterative executions and time intervals between each execution. $N_{repeat}$ controls how long the CPU frequency will stay in the increased state. Conversely, $T_i$ adjusts the time interval between two iterative executions that define the idle period of the runtime of this program. As the values of these two parameters are selected randomly during the noise injection process, the CPU frequency readings are also randomized. With the injected noise, the original frequency fingerprints of each docker image change randomly. We first tested the efficacy of our noise injection tool for the first scenario where the adversary does not know about the deployed tool. The pre-trained model (trained with the noiseless dataset) only achieves a 9\% detection accuracy rate when noisy fingerprints are classified.

In the second scenario, the adversary is aware of the noise injection tool. Hence, 
the attacker might collect noisy fingerprints and train a new  model with the injected noise to circumvent the effect of the noise injection tool. In this case, the detection accuracy for the new model is 50.2\% with the noisy dataset. This proves our proposed program can produce significant noise and degrade the performance of the deep-learning model trained with the noisy dataset. The performance overhead results are given in Appendix~\ref{appendix:6}.

%Noise injection is a known technique to mask the frequency fingerprint of containers in different platforms. %The noise injection process can be implemented by creating specially crafted code snippets to increase the frequency of the sibling cores at random times. The objective is to introduce the minimum amount of noise necessary to conceal the container's frequency signature while keeping the associated overhead low. 
%.  
%To generate noise, one could draw from the principles of differential privacy~\cite{dwork2006differential}, which can offer enhanced security guarantees. The random noise injection technique has been successful against website fingerprinting attacks by masking cache~\cite{shusterman2019robust} and interrupt~\cite{cook2022there} fingerprints. Since side-channel attacks mostly rely on machine learning models, it is even possible to inject targeted noise, leading to drastically changed fingerprints and even lower detection rates~\cite{gulmezoglu2021xai}. 
%Noise injection is a known defense technique to mask or alter fingerprints~\cite{dwork2006differential, shusterman2019robust, cook2022there, gulmezoglu2021xai}.

%based on noise injection, which significantly alters fingerprints to mask the rapid frequency changes. 

%Our defense mechanism injects random noise into the attacker's CPU frequency readings by increasing the workload in a controlled manner during the runtime of a docker image.

\smallskip
\noindent\textbf{Syscall Pattern Monitoring.}
To carry out the dynamic frequency attacks, adversaries need to repeatedly access the information supplied by \textit{cpufreq}. Given this insight, an alternative countermeasure could involve monitoring the system calls invoked to identify potential malicious activities. If a suspicious pattern is detected, the cloud admin can be notified, stating that a malicious user-space application is profiling running containers. One of the key advantages of this countermeasure is its cost-effectiveness, as there exist tools for efficiently monitoring system calls on Linux systems without adversely affecting the applications' performance. %While there are several tools to monitor the system calls on Linux systems, 
Using the \textit{perf trace} tool~\cite{perf}, we monitor the system calls invoked on the server used for the Firecracker and gVisor, both with and without an ongoing attack. First, the attack execution is monitored with the profiling tool to examine the \textit{syscall} patterns while accessing the \textit{cpufreq} interface. We observed four distinct system calls used to access the frequency values in order: fstat, fadvise64, read, and close. All these system calls have an input of the \textit{cpufreq} folder. For the benign process dataset collection, 50 Phoronix benchmark tests are selected randomly and monitored with the perf tool for a duration of 60 seconds. The syscall pattern we observed in the attack execution is not visible in the benchmark tests, resulting in 0\% false positive and 100\% attack detection rate. When we analyze the performance overhead of the perf tool with randomly selected 20 benchmark tests, we observe that the average performance overhead is 1.8\% ranging between 0.1\% and 8.6\% as shown in Appendix~\ref{appendix:5} \autoref{fig:perf_overhead}. Our detection tool shows that malicious applications exploiting the dynamic frequency channel can be detected with a high success rate and minimal performance overhead. Our preliminary experiments demonstrate that such a detection tool is a promising research direction to mitigate our proposed attack.

\smallskip
\noindent\textbf{Restricting Access Privilege to cpufreq.} Dynamic frequency attacks leverage the user-space access to the \textit{cpufreq} interface to access the current frequency values of each core. The simplest way of mitigating dynamic frequency attacks would be to restrict access to the interface for root privileged users only. However, this countermeasure is still ineffective against malicious (or honest-but-curious) entities with root privileges in the host machines, such as cloud providers. Hence, there is a need for user-controlled defense mechanisms to lower the attacker's capabilities.

\begin{comment}
\begin{figure}[t!]
    \centering
    \includegraphics[width=\linewidth]{Figures/perf_overhead.eps}
    \caption{Performance overhead for the 20 Phoronix benchmark tests. The overall performance overhead is 1.8\%}
    \label{fig:perf_overhead}
\end{figure}
\end{comment}

\begin{comment}
\begin{algorithm}[ht!]
\small
\caption{\textcolor{blue}{Pseudo code of noise injection}\label{alg:noise}}
%\tcp{$repeat$ Pointer to the address of the buffer end}

\tcp{$N_{repeat}$ Number of repeated execution}
\tcp{$T_{sleep}$ Time interval between two consecutive repetitions}

\KwInput{$N_{repeat}, T_{i}$}

\textcolor{blue}{define} variables $a,b,c$ \\
%\tcp{Running SMC for $N_{repeat}$ times with $T_{sleep}$ $\mu s$ interval}
\For{$i \gets 1$ to $N_{repeat}$}
{
    \textbf{\_\_asm\_\_}\\
    \{\\
          \quad \textcolor{blue}{fld} \: $\%1\backslash n$ \\
          \quad \textcolor{blue}{fld} \: $\%2\backslash n$  \\
          \quad \textcolor{blue}{faddp}$\backslash n$\\
          \quad \textcolor{blue}{fmulp}$\backslash n$\\
          \quad \textcolor{blue}{fstp}  \: $\%0\backslash n$ \\
          \quad : "=m" (c) \\
          \quad : "m" (a), "m" (b) \\
     \}
}
\end{algorithm}
\end{comment}

\begin{comment}
\begin{figure}[t!]
    \centering
    \includegraphics[width=\linewidth]{Figures/overhead.png}
    \caption{\textcolor{red}{The overhead in terms of system-wide energy consumption calculated over 25 different images.}}
    \label{fig:overhead_noise_inj}
\end{figure}
\end{comment}
\section{Conclusion}\label{sec:conclusion}

This study analyzes the possibility of identifying running containers through dynamic frequency fingerprints in native, sandboxed, and TEEs. We show that our attack achieves at least a 70\% success rate in all these environments. We examine various scenarios that an attacker can face in cloud computing, such as multiple core utilization, the influence of Docker image pull and MicroVM bootstrapping processes, a diverse set of input arguments, and the detection of different image versions. We also perform our attack on a diverse set of microarchitectures, as well as running multiple containers simultaneously, and classify them with high accuracy. Finally, we propose two countermeasures: (1) a client-based defense technique using artificial noise injection to reduce the detection accuracy and (2) a lightweight cloud-based countermeasure to profile system calls in a system and distinguish malicious activities with only 1.8\% performance overhead while achieving a 100\% detection rate. This study opens new research directions in the security of both sandbox and TEE applications in cloud computing. 

% \ifthenelse{\thepage > 12}{{\Huge\bfseries\color{red} Over page limit! \thepage{} > 12}}{}

\section*{Acknowledgements}
We thank the anonymous reviewers for their insightful feedback and help in improving this paper. This work was partially funded by the German Research Foundation (DFG) under grant no. 439797619 and 456967092. Moreover, the research leading to these results have received funding from the European Union’s Horizon 2020 research and innovation programme under grant agreements No 101070473 (FLUIDOS) and No 101092950 (EDGELESS) and from the UNICO I+D Cloud programme with the CLOUDLESS project.

% the "some research foundation" under grant no. 111111 and 222222 and the "some other foundation" under grant no. 333333 and 444444.

%\newpage
\bibliographystyle{plain}
\bibliography{Ref}

\begin{thebibliography}{10}

\bibitem{cve-2017-5123}
{epanaroma: Escape Docker Container Using waitid() | CVE-2017-5123 | Twistlock}.
\newblock \url{https://www.epanorama.net/blog/2018/01/05/escape-docker-container-using-waitid-cve-2017-5123-twistlock-19/}, 2017.

\bibitem{aciiccmez2007yet}
Onur Acii{\c{c}}mez.
\newblock Yet another microarchitectural attack: : exploiting i-cache.
\newblock In {\em {CSAW}}, pages 11--18. {ACM}, 2007.

\bibitem{aciiccmez2007power}
Onur Acii{\c{c}}mez, {\c{C}}etin~Kaya Ko{\c{c}}, and Jean-Pierre Seifert.
\newblock On the power of simple branch prediction analysis.
\newblock In {\em AsiaCCS}, pages 312--320. {ACM}, 2007.

\bibitem{agache2020firecracker}
Alexandru Agache, Marc Brooker, Alexandra Iordache, Anthony Liguori, Rolf Neugebauer, Phil Piwonka, and Diana{-}Maria Popa.
\newblock Firecracker: Lightweight virtualization for serverless applications.
\newblock In {\em {NSDI}}, pages 419--434. {USENIX} Association, 2020.

\bibitem{aldaya2019port}
Alejandro~Cabrera Aldaya, Billy~Bob Brumley, Sohaib ul~Hassan, Cesar~Pereida Garc{\'{\i}}a, and Nicola Tuveri.
\newblock Port contention for fun and profit.
\newblock In {\em {IEEE} Symposium on Security and Privacy}, pages 870--887. {IEEE}, 2019.

\bibitem{amd_turbo}
Turbo core technology, 2023.
\newblock \url{https://www.amd.com/en/technologies/turbo-core}.

\bibitem{amd_sev}
AMD.
\newblock {Google Cloud Confidential Computing Powered by AMD}, Last accessed: 03-17-2024.
\newblock \url{https://www.amd.com/en/solutions/google-cloud-confidential-computing}.

\bibitem{irazoqui2014wait}
Gorka~Irazoqui Apecechea, Mehmet~Sinan Inci, Thomas Eisenbarth, and Berk Sunar.
\newblock Wait a minute! {A} fast, cross-vm attack on {AES}.
\newblock In {\em {RAID}}, volume 8688 of {\em Lecture Notes in Computer Science}, pages 299--319. Springer, 2014.

\bibitem{Apparmor}
Apparmor -- linux kernel security module, 2019.
\newblock \url{https://apparmor.net/}.

\bibitem{arnautov2016scone}
Sergei Arnautov, Bohdan Trach, Franz Gregor, Thomas Knauth, Andr{\'{e}} Martin, Christian Priebe, Joshua Lind, Divya Muthukumaran, Dan O'Keeffe, Mark Stillwell, David Goltzsche, David~M. Eyers, R{\"{u}}diger Kapitza, Peter~R. Pietzuch, and Christof Fetzer.
\newblock {SCONE:} secure linux containers with intel {SGX}.
\newblock In {\em {OSDI}}, pages 689--703. {USENIX} Association, 2016.

\bibitem{aws2023nitrosecurity}
J.~D. Bean, Mark Ryland, Matthew~S. Wilson, Colm MacCárthaigh, and Benjamin Serebrin.
\newblock The security design of the aws nitro system.
\newblock White paper, AWS, Nov 2022.

\bibitem{van2019tale}
Jo~Van Bulck, David~F. Oswald, Eduard Marin, Abdulla Aldoseri, Flavio~D. Garcia, and Frank Piessens.
\newblock A tale of two worlds: Assessing the vulnerability of enclave shielding runtimes.
\newblock In {\em {CCS}}, pages 1741--1758. {ACM}, 2019.

\bibitem{alibaba_ecs}
Alibaba Cloud.
\newblock {Elastic Compute Service}, Last accessed: 03-17-2024.
\newblock \url{https://www.alibabacloud.com/help/en/ecs/user-guide/step-1-deploy-a-client}.

\bibitem{amd_google_cloud}
Google Cloud.
\newblock {AMD and Google Cloud}, Last accessed: 03-17-2024.
\newblock \url{https://cloud.google.com/amd}.

\bibitem{7742298}
Th{\'{e}}o Combe, Antony Martin, and Roberto~Di Pietro.
\newblock To docker or not to docker: {A} security perspective.
\newblock {\em {IEEE} Cloud Comput.}, 3(5):54--62, 2016.

\bibitem{constable2023aexnotify}
Scott Constable, Jo~Van~Bulck, Xiang Cheng, Yuan Xiao, Cedric Xing, Ilya Alexandrovich, Taesoo Kim, Frank Piessens, Mona Vij, and Mark Silberstein.
\newblock Aex-notify: Thwarting precise single-stepping attacks through interrupt awareness for intel sgx enclaves.
\newblock In {\em 32nd {USENIX} Security Symposium}, pages 4051--4068, August 2023.

\bibitem{containerd}
containerd, Last accessed: 08-21-2023.
\newblock \url{https://containerd.io/}.

\bibitem{costan2016intel}
Victor Costan and Srinivas Devadas.
\newblock Intel {SGX} explained.
\newblock {\em {IACR} Cryptol. ePrint Arch.}, page~86, 2016.

\bibitem{delimitrou2017bolt}
Christina Delimitrou and Christos Kozyrakis.
\newblock Bolt: {I} know what you did last summer... in the cloud.
\newblock In {\em {ASPLOS}}, pages 599--613. {ACM}, 2017.

\bibitem{dipta2022df}
Debopriya~Roy Dipta and Berk G{\"{u}}lmezoglu.
\newblock {DF-SCA:} dynamic frequency side channel attacks are practical.
\newblock In {\em {ACSAC}}, pages 841--853. {ACM}, 2022.

\bibitem{DockerHubAPI}
Docker hub api, 2023.
\newblock \url{https://docs.docker.com/docker-hub/api/latest/}.

\bibitem{Seccomp}
Seccomp security profiles for docker, 2023.
\newblock \url{https://docs.docker.com/engine/security/seccomp/}.

\bibitem{docker}
Use containers to build, share and run your applications, 2023.
\newblock \url{https://www.docker.com/resources/what-container/}.

\bibitem{DBLP:conf/ndss/FangWNOSKRH22}
Chongzhou Fang, Han Wang, Najmeh Nazari, Behnam Omidi, Avesta Sasan, Khaled~N. Khasawneh, Setareh Rafatirad, and Houman Homayoun.
\newblock {Repttack: Exploiting Cloud Schedulers to Guide Co-Location Attacks}.
\newblock In {\em Annual Network and Distributed System Security Symposium, {NDSS}}, 2022.

\bibitem{gao2017containerleaks}
Xing Gao, Zhongshu Gu, Mehmet Kayaalp, Dimitrios Pendarakis, and Haining Wang.
\newblock Containerleaks: Emerging security threats of information leakages in container clouds.
\newblock In {\em {DSN}}, pages 237--248. {IEEE} Computer Society, 2017.

\bibitem{gramine}
Gramine - a library os for unmodified applications, 2023.
\newblock \url{https://gramineproject.io/}.

\bibitem{gsc}
gsc – gramine shielded containers, 2023.
\newblock \url{https://gramine.readthedocs.io/projects/gsc/en/latest/}.

\bibitem{gulmezoglu2017cache}
Berk G{\"{u}}lmezoglu, Thomas Eisenbarth, and Berk Sunar.
\newblock Cache-based application detection in the cloud using machine learning.
\newblock In {\em AsiaCCS}, pages 288--300. {ACM}, 2017.

\bibitem{gulmezoglu2017perfweb}
Berk G{\"{u}}lmezoglu, Andreas Zankl, Thomas Eisenbarth, and Berk Sunar.
\newblock Perfweb: How to violate web privacy with hardware performance events.
\newblock In {\em {ESORICS} {(2)}}, volume 10493 of {\em Lecture Notes in Computer Science}, pages 80--97. Springer, 2017.

\bibitem{inci2015seriously}
Mehmet~Sinan Inci, Berk G{\"{u}}lmezoglu, Gorka~Irazoqui Apecechea, Thomas Eisenbarth, and Berk Sunar.
\newblock Seriously, get off my cloud! cross-vm {RSA} key recovery in a public cloud.
\newblock {\em {IACR} Cryptol. ePrint Arch.}, page 898, 2015.

\bibitem{inci2016co}
Mehmet~Sinan Inci, Berk G{\"{u}}lmezoglu, Thomas Eisenbarth, and Berk Sunar.
\newblock Co-location detection on the cloud.
\newblock In {\em {COSADE}}, volume 9689 of {\em Lecture Notes in Computer Science}, pages 19--34. Springer, 2016.

\bibitem{inci2016cache}
Mehmet~Sinan Inci, Berk G{\"{u}}lmezoglu, Gorka Irazoqui, Thomas Eisenbarth, and Berk Sunar.
\newblock Cache attacks enable bulk key recovery on the cloud.
\newblock In {\em {CHES}}, volume 9813 of {\em Lecture Notes in Computer Science}, pages 368--388. Springer, 2016.

\bibitem{intel_turbo}
What is intel turbo boost technology?, 2023.
\newblock \url{https://www.intel.com/content/www/us/en/gaming/resources/turbo-boost.html}.

\bibitem{cryptoeprint:2016/086}
{Intel SGX Explained}.
\newblock Cryptology ePrint Archive, Paper 2016/086, 2016.
\newblock \url{https://eprint.iacr.org/2016/086}.

\bibitem{sev2017seves}
David Kaplan.
\newblock Protecting {VM} register state with {SEV-ES}.
\newblock White paper, AMD, Feb 2017.

\bibitem{firecracker2023buildfcvm}
Kazuyoshi Kato, Maksym Pavlenko, Erik Sipsma, Samuel Karp, xibz, Austin Vazquez, Kern Walster, Gavin Inglis, Noah Meyerhans, Anirudh Aithal, Alakesh Haloi, et~al.
\newblock Create firecracker {VM} images for use with firecracker-containerd, 2023.

\bibitem{perf}
Michael Kerrisk.
\newblock perf-trace(1) -- linux manual page, 2023.
\newblock \url{https://man7.org/linux/man-pages/man1/perf-trace.1.html}.

\bibitem{kim2022thermalbleed}
Taehun Kim and Youngjoo Shin.
\newblock Thermalbleed: {A} practical thermal side-channel attack.
\newblock {\em {IEEE} Access}, 10:25718--25731, 2022.

\bibitem{lipp2021platypus}
Moritz Lipp, Andreas Kogler, David~F. Oswald, Michael Schwarz, Catherine Easdon, Claudio Canella, and Daniel Gruss.
\newblock {PLATYPUS:} software-based power side-channel attacks on x86.
\newblock In {\em {IEEE} Symposium on Security and Privacy}, pages 355--371. {IEEE}, 2021.

\bibitem{7807249}
Pieter Maene, Johannes Götzfried, Ruan de~Clercq, Tilo Müller, Felix Freiling, and Ingrid Verbauwhede.
\newblock Hardware-based trusted computing architectures for isolation and attestation.
\newblock {\em IEEE Transactions on Computers}, 67(3):361--374, 2018.

\bibitem{moghimi2017cachezoom}
Ahmad Moghimi, Gorka Irazoqui, and Thomas Eisenbarth.
\newblock Cachezoom: How {SGX} amplifies the power of cache attacks.
\newblock In {\em {CHES}}, volume 10529 of {\em Lecture Notes in Computer Science}, pages 69--90. Springer, 2017.

\bibitem{naghibijouybari2018rendered}
Hoda Naghibijouybari, Ajaya Neupane, Zhiyun Qian, and Nael~B. Abu{-}Ghazaleh.
\newblock Rendered insecure: {GPU} side channel attacks are practical.
\newblock In {\em {CCS}}, pages 2139--2153. {ACM}, 2018.

\bibitem{cve6662}
NIST.
\newblock {CVE-2016-6662}, Last accessed: 11-2-2023.
\newblock \url{https://nvd.nist.gov/vuln/detail/CVE-2016-6662}.

\bibitem{cve27928}
NIST.
\newblock {CVE-2021-27928}, Last accessed: 11-2-2023.
\newblock \url{https://nvd.nist.gov/vuln/detail/CVE-2021-27928}.

\bibitem{paccagnella2021lord}
Riccardo Paccagnella, Licheng Luo, and Christopher~W. Fletcher.
\newblock Lord of the ring(s): Side channel attacks on the {CPU} on-chip ring interconnect are practical.
\newblock In {\em {USENIX} Security Symposium}, pages 645--662. {USENIX} Association, 2021.

\bibitem{Selinux}
What is selinux, 2019.
\newblock \url{https://www.redhat.com/en/topics/linux/what-is-selinux}.

\bibitem{ristenpart2009hey}
Thomas Ristenpart, Eran Tromer, Hovav Shacham, and Stefan Savage.
\newblock Hey, you, get off of my cloud: exploring information leakage in third-party compute clouds.
\newblock In {\em {CCS}}, pages 199--212. {ACM}, 2009.

\bibitem{salmen2021containervmconversion}
Kai Salmen.
\newblock Automatic conversion of containers to virtual machines, 2021.
\newblock \url{https://www.typefox.io/blog/automatic-conversion-of-containers-to-virtual-machines/}.

\bibitem{aws_amd}
Amazon~Web Services.
\newblock {AMD SEV-SNP}, Last accessed: 03-17-2024.
\newblock \url{https://docs.aws.amazon.com/AWSEC2/latest/UserGuide/sev-snp.html}.

\bibitem{aws_instance}
Amazon~Web Services.
\newblock {AWS and Intel}, Last accessed: 03-17-2024.
\newblock \url{https://aws.amazon.com/intel/}.

\bibitem{taneja2023hot}
Hritvik Taneja, Jason Kim, Jie~Jeff Xu, Stephan van Schaik, Daniel Genkin, and Yuval Yarom.
\newblock Hot pixels: Frequency, power, and temperature attacks on gpus and arm socs.
\newblock In {\em {USENIX} Security Symposium}, pages 6275--6292. {USENIX} Association, 2023.

\bibitem{tsai2017graphene}
Chia{-}che Tsai, Donald~E. Porter, and Mona Vij.
\newblock Graphene-sgx: {A} practical library {OS} for unmodified applications on {SGX}.
\newblock In {\em {USENIX} Annual Technical Conference}, pages 645--658. {USENIX} Association, 2017.

\bibitem{vanbulck2017sgxstep}
Jo~Van~Bulck, Frank Piessens, and Raoul Strackx.
\newblock {SGX-Step}: A practical attack framework for precise enclave execution control.
\newblock In {\em 2nd Workshop on System Software for Trusted Execution {(SysTEX)}}, pages 4:1--4:6. {ACM}, October 2017.

\bibitem{gvisor}
Fabricio Voznika, Adin Scannell, Andrei Vagin, Kevin Krakauer, Ayush Ranjan, Nicolas Lacasse, Ghanan, Jamie Liu, Michael Pratt, Bhasker Hariharan, et~al.
\newblock gvisor, 2023.

\bibitem{wang2018peeking}
Liang Wang, Mengyuan Li, Yinqian Zhang, Thomas Ristenpart, and Michael~M. Swift.
\newblock Peeking behind the curtains of serverless platforms.
\newblock In {\em {USENIX} Annual Technical Conference}, pages 133--146. {USENIX} Association, 2018.

\bibitem{wang2017leaky}
Wenhao Wang, Guoxing Chen, Xiaorui Pan, Yinqian Zhang, XiaoFeng Wang, Vincent Bindschaedler, Haixu Tang, and Carl~A. Gunter.
\newblock Leaky cauldron on the dark land: Understanding memory side-channel hazards in {SGX}.
\newblock In {\em {CCS}}, pages 2421--2434. {ACM}, 2017.

\bibitem{wang2022hertzbleed}
Yingchen Wang, Riccardo Paccagnella, Elizabeth~Tang He, Hovav Shacham, Christopher~W. Fletcher, and David Kohlbrenner.
\newblock Hertzbleed: Turning power side-channel attacks into remote timing attacks on x86.
\newblock In {\em {USENIX} Security Symposium}, pages 679--697. {USENIX} Association, 2022.

\bibitem{kata2023containervmconversion}
Li~Wei, Jonathan Bryce, Xu~Wang, Peng Tao, Fabiano Fidêncio, Hui Zhu, Mohammed Naser, Anastassios Nanos, Thierry Carrez, Maksym Pavlenko, et~al.
\newblock Kata containers architecture -- container creation, 2023.

\bibitem{wu2014whispers}
Zhenyu Wu, Zhang Xu, and Haining Wang.
\newblock Whispers in the hyper-space: High-bandwidth and reliable covert channel attacks inside the cloud.
\newblock {\em {IEEE/ACM} Trans. Netw.}, 23(2):603--615, 2015.

\bibitem{young2019true}
Ethan~G. Young, Pengfei Zhu, Tyler Caraza{-}Harter, Andrea~C. Arpaci{-}Dusseau, and Remzi~H. Arpaci{-}Dusseau.
\newblock The true cost of containing: {A} gvisor case study.
\newblock In {\em HotCloud}. {USENIX} Association, 2019.

\bibitem{zhang2009peeping}
Kehuan Zhang and XiaoFeng Wang.
\newblock Peeping tom in the neighborhood: Keystroke eavesdropping on multi-user systems.
\newblock In {\em {USENIX} Security Symposium}, pages 17--32. {USENIX} Association, 2009.

\bibitem{zhang2012cross}
Yinqian Zhang, Ari Juels, Michael~K. Reiter, and Thomas Ristenpart.
\newblock Cross-vm side channels and their use to extract private keys.
\newblock In {\em {CCS}}, pages 305--316. {ACM}, 2012.

\bibitem{zhang2014cross}
Yinqian Zhang, Ari Juels, Michael~K. Reiter, and Thomas Ristenpart.
\newblock Cross-tenant side-channel attacks in paas clouds.
\newblock In {\em {CCS}}, pages 990--1003. {ACM}, 2014.

\bibitem{zhang2021red}
Zhenkai Zhang, Sisheng Liang, Fan Yao, and Xing Gao.
\newblock Red alert for power leakage: Exploiting intel rapl-induced side channels.
\newblock In {\em AsiaCCS}, pages 162--175. {ACM}, 2021.

\end{thebibliography}

\appendices

\newpage
\section{Impact of Running Multiple Containers on the Same Thread.} \label{appendix:1}
Although it is not ideal to run multiple containers on the same thread of a core, we analyze the impact of our attack in this threat scenario. For this analysis, we randomly chose ten containers and ran a pair of containers concurrently on the same thread of a core. Considering two different containers run simultaneously, we have 45 different possible combinations of pairs made from the list of 10 containers that we chose previously. We collected fingerprints for all of these combinations. The intention is to analyze whether it is possible to recognize the specific combination or the pair of containers by looking at the combined frequency fingerprint if two containers run at a time in the same thread. To verify that concept, we trained a model with frequency fingerprints of all these 45 combinations and tested the accuracy. The accuracy of the model to correctly predict the pair of containers is 64.67\%. Considering only ten containers, the accuracy shows that it is challenging to predict the correct pair of containers if they run in the same thread as fingerprints are combined together in one fingerprint, which degrades the model accuracy. 

\section{Noise signature of Firecracker VMM.}
\label{appendix:2}

\newpage

\section{Frequency fingerprints of pulling images inside MicroVM.}
\label{appendix:3}

\begin{figure}[ht!]
    \centering
    \includegraphics[width=\linewidth]{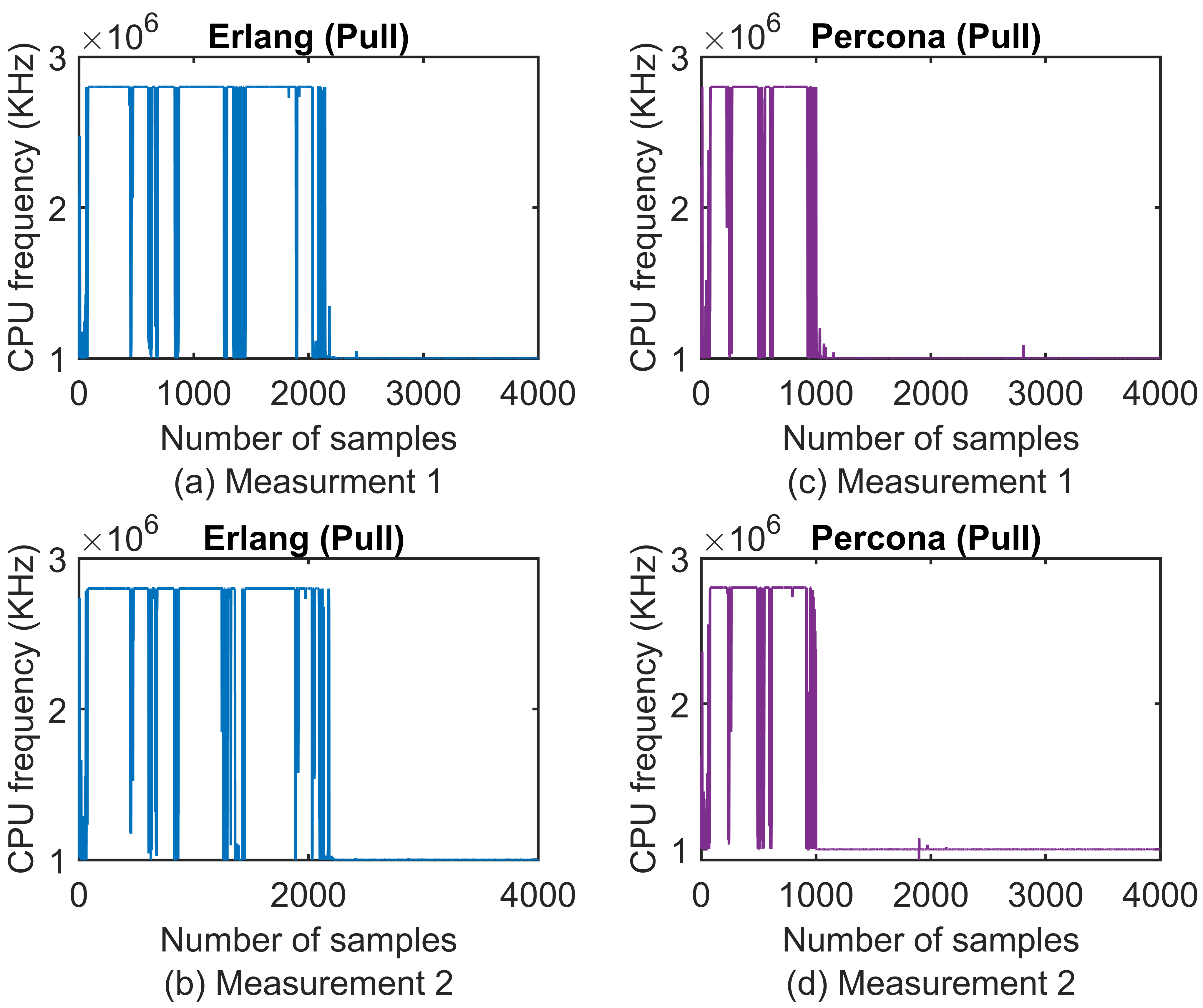}
    \caption{Comparison of pulling different images inside a MicroVM of Firecracker. (a, b) represent two different measurements of pulling the \texttt{Erlang} image, and (c, d) refers to two separate measurements for pulling the \texttt{Percona} image. As can be seen, it is possible to recognize the container image being pulled in the Firecracker execution environment based on the CPU frequency signatures.}
    \label{fig:pull}
\end{figure}

\section{ Configuration Script for Firecracker.}
\label{appendix:4}

\begin{itemize}[leftmargin=*,noitemsep]
  \item \textbf{Disable plugins:} The plugins are simply extensions that are supposed to enhance specific \texttt{firecracker-containerd} capabilities. The \texttt{io.containerd.grpc.v1.cri} plugin is disabled through the configuration file, i.\,e., this certain container runtime feature is excluded from the setup.

  \item \textbf{Storage and State Management:} The configuration file defines the root directory for firecracker-containerd's storage. The state information of the firecracker-containerd is also stored in a separate directory specified by the configuration file.

  \item \textbf{Configuration of gRPC setting:} gRPC is a high-level communication protocol that ensures communications between services. 
  One of the best capabilities of gRPC is its multiplexing capability that works across different languages and platforms. In the configuration file, the address of the Unix domain socket is specified to ensure seamless communication for gRPC.

  \item \textbf{Configuration of snapshot plugin:} The \texttt{io.containerd.snapshotter.v1.devmapper} plugin is configured to utilize the Devmapper technology. This plugin specifically helps to manage the container images effectively. To configure this setting, a storage space of 10\,GB is allocated for the efficient management of the container images. 

  \item \textbf{Debug capability:} The debug capability is added through the configuration file by setting the log level which gives an opportunity for troubleshooting.
  
\end{itemize}

\section{Performance Overhead for Syscall Pattern Monitoring.}
\label{appendix:5}

\begin{figure}[ht!]
    \centering
    \includegraphics[width=\linewidth]{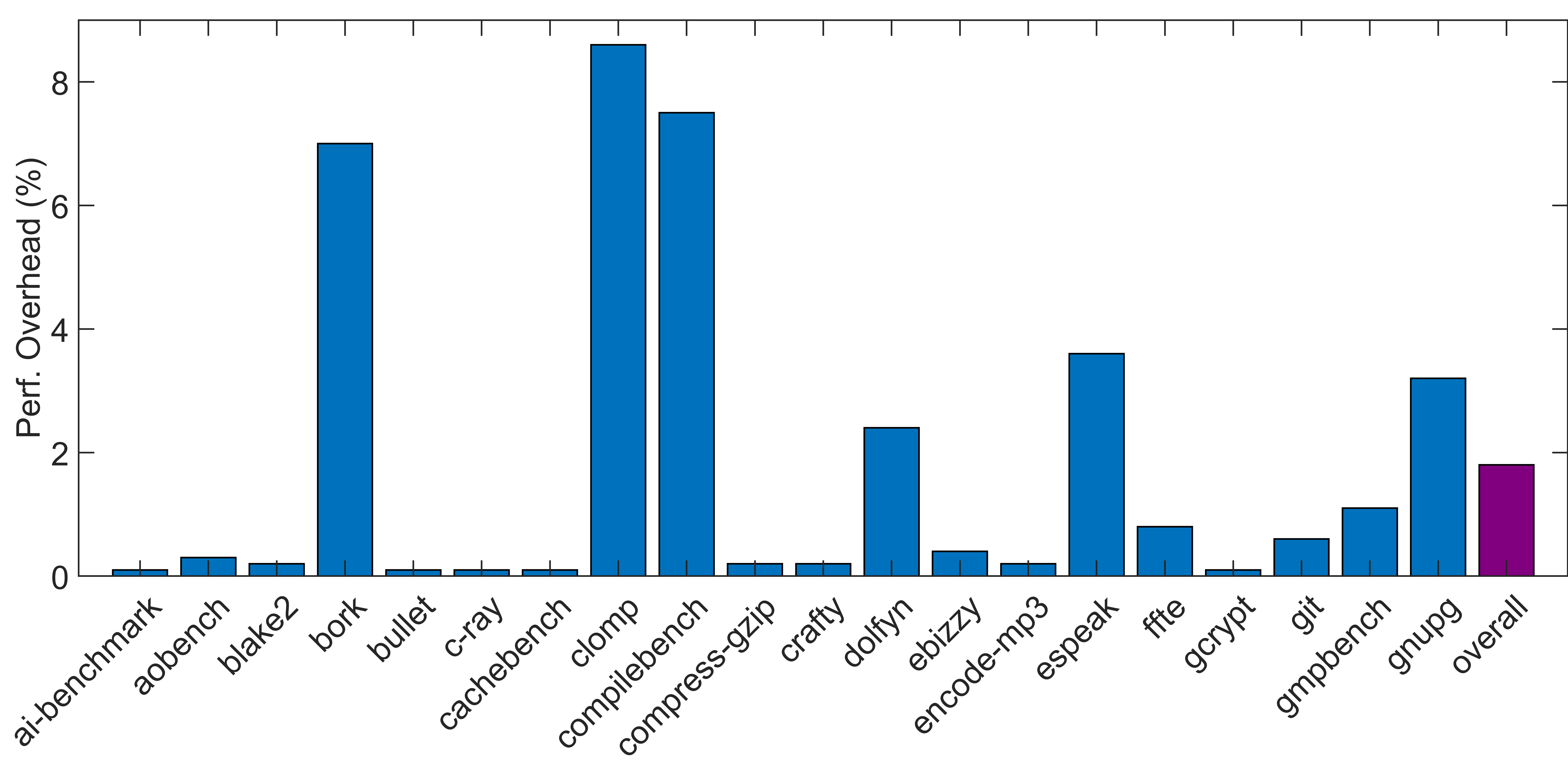}
    \caption{Performance overhead for the 20 Phoronix benchmark tests. The overall performance overhead is 1.8\%}
    \label{fig:perf_overhead}
\end{figure}

\section{Performance Overhead for the Noise Injection Tool.}
\label{appendix:6}

\begin{figure}[ht!]
    \centering
    \includegraphics[width=\linewidth]{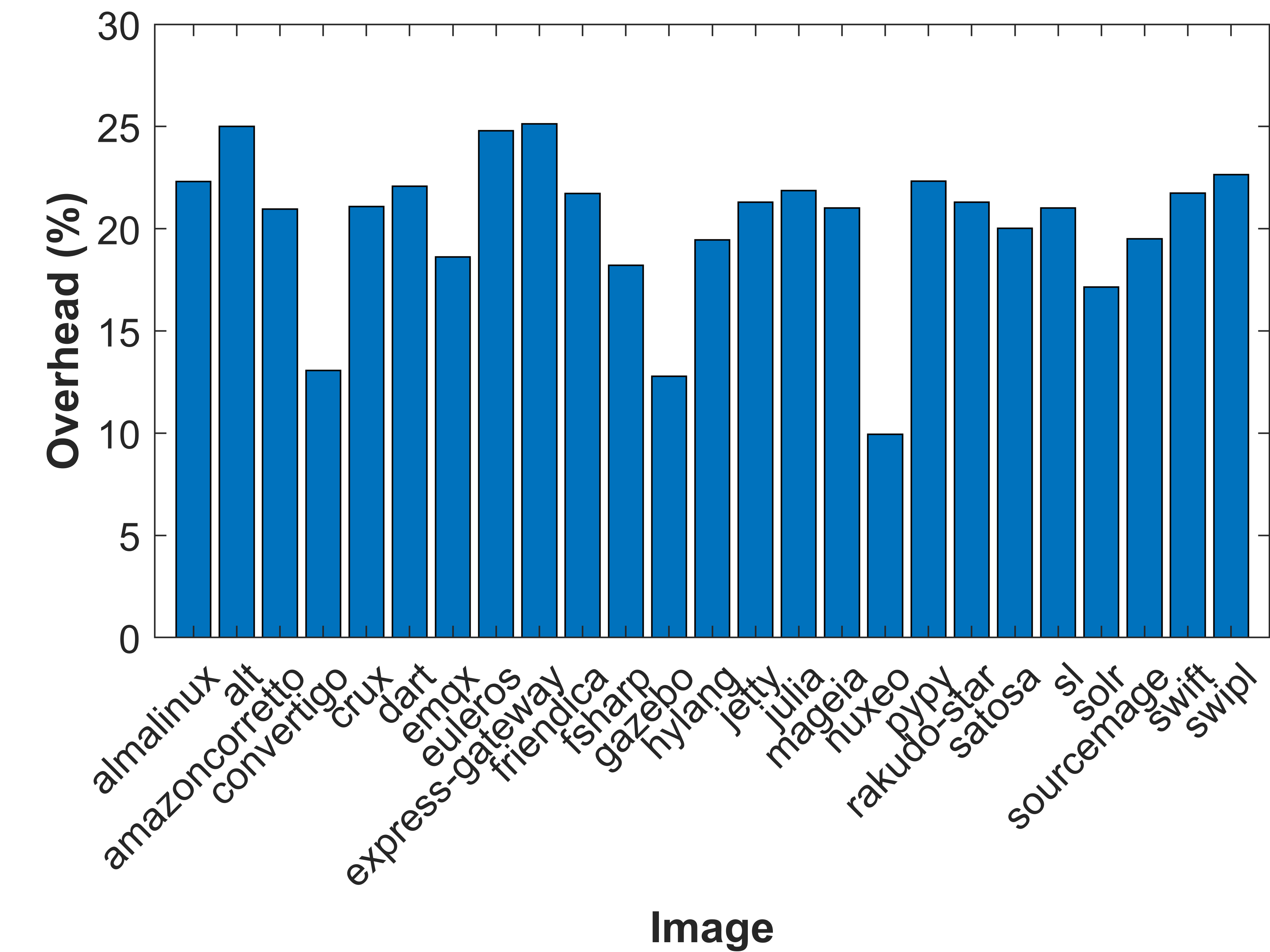}
    \caption{The overhead in terms of system-wide energy consumption calculated over 25 different images.}
    \label{fig:overhead_noise_inj}
\end{figure}

As the proposed noise injection program will run in the background, it is important to calculate the performance overhead in terms of energy consumption. For measuring the system-wide energy consumption, we leveraged the Intel RAPL tool. We calculated the performance overhead for 25 images randomly picked from our image list. In Figure~\ref{fig:overhead_noise_inj}, we show that our noise injection tool introduces 19.7\% overhead on average with a maximum of 25.1\% performance overhead.

\newpage
\section{Pseudo code of the Noise Injection Tool.}
\label{appendix:7}
\begin{algorithm}[ht!]
\small
\caption{\textcolor{blue}{Pseudo code of noise injection}\label{alg:noise}}
%\tcp{$repeat$ Pointer to the address of the buffer end}

\tcp{$N_{repeat}$ Number of repeated execution}
\tcp{$T_{sleep}$ Time interval between two consecutive repetitions}

\KwInput{$N_{repeat}, T_{i}$}

\textcolor{blue}{define} variables $a,b,c$ \\
%\tcp{Running SMC for $N_{repeat}$ times with $T_{sleep}$ $\mu s$ interval}
\For{$i \gets 1$ to $N_{repeat}$}
{
    \textbf{\_\_asm\_\_}\\
    \{\\
          \quad \textcolor{blue}{fld} \: $\%1\backslash n$ \\
          \quad \textcolor{blue}{fld} \: $\%2\backslash n$  \\
          \quad \textcolor{blue}{faddp}$\backslash n$\\
          \quad \textcolor{blue}{fmulp}$\backslash n$\\
          \quad \textcolor{blue}{fstp}  \: $\%0\backslash n$ \\
          \quad : "=m" (c) \\
          \quad : "m" (a), "m" (b) \\
     \}
}
\end{algorithm}

\section{List of input arguments to run the container.}
\label{appendix:8}

\begin{tabular}{p{.325\linewidth}p{.5\linewidth}}
    \toprule
    \bfseries Argument & \bfseries Description \\
    \midrule
    \multicolumn{2}{l}{\emph{minimal arguments}} \\
    \midrule
    \ttfamily name         & Assign a name to the container \\ \addlinespace
    \ttfamily rm           & Automatically remove the container when it exits \\ \addlinespace
    \ttfamily cpuset-cpus  & CPUs in which to allow execution \\ \addlinespace
    \ttfamily it           & Assign a pseudo-TTY to the container  \\ \addlinespace
    \ttfamily runtime      & Container runtime specific to the sandbox environment \\
    \midrule
    \multicolumn{2}{l}{\emph{additional arguments}} \\
    \midrule
    \ttfamily v            & Create a bind mount \\ \addlinespace
    \ttfamily p            & Publish a container's port, or range of ports, to the host \\ \addlinespace
    \ttfamily e            & Set environment variables \\ \addlinespace
    \ttfamily input script & Configuration file or script to be executed (Specific to an image) \\
    \bottomrule
\end{tabular}



\end{document}